\documentclass{aa}
\AddToHook{begindocument/before}{\usepackage{hyperref}}
\usepackage{graphicx}
\usepackage{txfonts}
\usepackage{color}
\usepackage{xcolor}
\usepackage{hyperref}
\usepackage{ulem}
\usepackage{sidecap}

\definecolor{dg}{rgb}{0.0, 0.5, 0.0}
\definecolor{applegreen}{rgb}{0.55, 0.71, 0.0}
\definecolor{dr}{rgb}{0.8, 0.2, 0.0}
\definecolor{orange}{rgb}{0.9, 0.5, 0.0}
\definecolor{db}{rgb}{0.00, 0.35, 0.80}
\definecolor{darkviolet}{rgb}{0.58, 0.0, 0.83}

\begin{document}

   \title{Flare heating of the chromosphere: \\ Observations of flare continuum from GREGOR and IRIS}
%   \subtitle{I. Overviewing the $\kappa$-mechanism}

   \author{M. Garc\'{i}a-Rivas\inst{1,2}
          \and J. Ka\v{s}parov\'{a}\inst{1}
          \and A. Berlicki\inst{1,3}
          \and M. \v{S}vanda\inst{1,2}
          \and J. Dud\'{i}k\inst{1}
          \and D. \v{C}tvrte\v{c}ka\inst{4}
          \and M. Zapi\'{o}r\inst{1}
          \and W. Liu\inst{1}
          \and M. Sobotka\inst{1}
          \and M. Pavelkov\'{a}\inst{1}
          \and G.~G. Motorina\inst{1} %\fmsep 
          \thanks{Currently at Central Astronomical Observatory at Pulkovo of Russian Academy of Sciences, St. Petersburg, 196140, Russia}
          }

   \institute{Astronomical Institute of the Czech Academy of Sciences, Fri\v{c}ova 298, 251\,65 Ond\v{r}ejov, Czech Republic\\
              \email{marta.garcia.rivas@asu.cas.cz}
        \and Astronomical Institute, Faculty of Mathematics and Physics, Charles University, V Hole\v{s}ovi\v{c}k\'ach 2, 182\,00 Prague, Czech Republic 
        \and Center of Scientific Excellence - Solar and Stellar Activity, University of Wrocław, Mikołaja Kopernika 11, 51-622 Wrocław, Poland 
        \and Christian Doppler Grammar School, Zborovsk\'a 621/45, 150\,00 Prague
        %\and {\color{green} Central Astronomical Observatory at Pulkovo of Russian Academy of Sciences, St. Petersburg, 196140, Russia}
        %\and {\color{red} Leibniz Institute for Astrophysics Potsdam (AIP), Potsdam, Germany}
        }

   \date{}

  \abstract
  % context heading (optional)
  {On 2022 May 4, an M5.7 flare erupted in the active region NOAA~13004, which was the target of a coordinated campaign between GREGOR, IRIS, Hinode, and ground-based instruments at the Ond\v{r}ejov observatory. A flare kernel located at the edge of a pore was co-observed by the IRIS slit and GREGOR HiFI+ imagers.  }
  % aims heading (mandatory)  
   {We investigated the flare continuum enhancement at different wavelength ranges in order to derive the temperature of the chromospheric layer heated during the flare. }
  % methods heading (mandatory)
   {All datasets were aligned to IRIS slit-jaw images. We selected a pixel along the IRIS slit where the flare kernel was captured and evaluated multi-wavelength light curves within it. We defined a narrow IRIS near-UV band that comprises only continuum emission. The method, which assumes that the flare continuum enhancement is due to optically thin emission from hydrogen recombination processes, was applied to obtain a lower limit on the temperature in the layer where the continuum enhancement was formed. }
  % results heading (mandatory)
   {We determined a lower limit for the temperature and its time evolution in the chromospheric layer heated during the flare in the range of (3 - 15)$\times10^3$\,K. The mean electron density in that layer was estimated to be $\sim 1\times10^{13}$~cm${}^{-3}$. } 
  % conclusions heading (optional)
   {Multi-wavelength flare co-observations are a rich source of diagnostics. Due to the rapidly evolving nature of flares, the sit-and-stare mode is key to achieving a high temporal cadence that allows one to thoroughly analyse the same flare structure. }


   \maketitle

   \nolinenumbers
%
%________________________________________________________________
\section{Introduction}
\label{Sect:Intro}

%
% FIGURE 1 - FLARE OVERVIEW
%--------------------------
\begin{figure*}
    \centering
    \includegraphics[width=17.6cm]{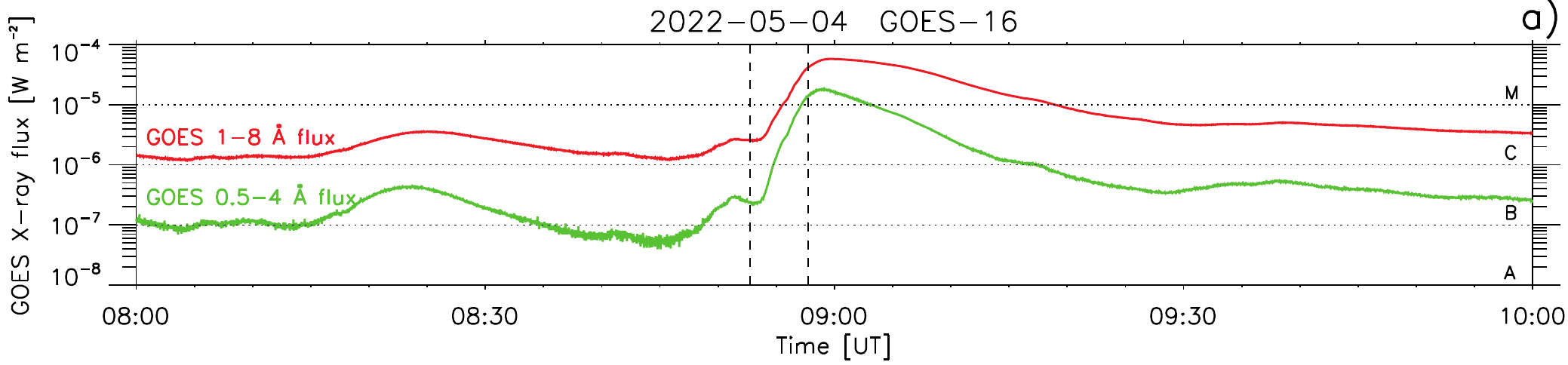 }
    \includegraphics[width=6.74cm, viewport= 0 40 334 225, clip]{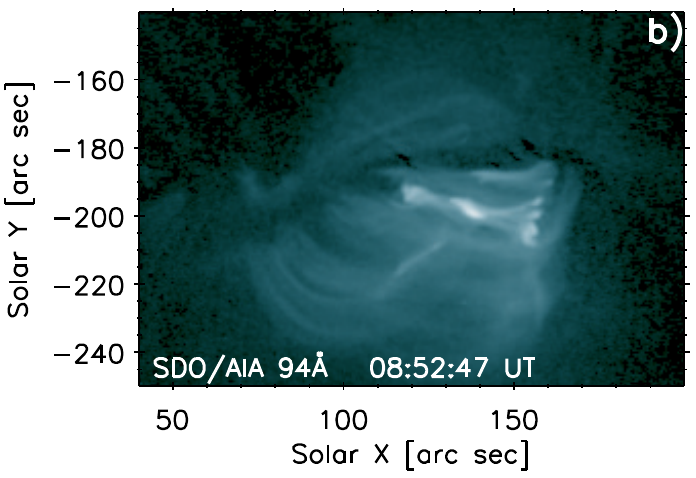 }
    \includegraphics[width=5.43cm, viewport=65 40 334 225, clip]{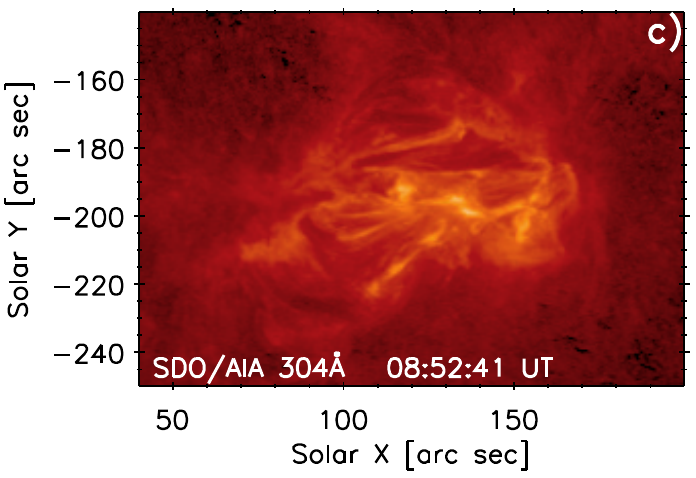 }
    \includegraphics[width=5.43cm, viewport=65 40 334 225, clip]{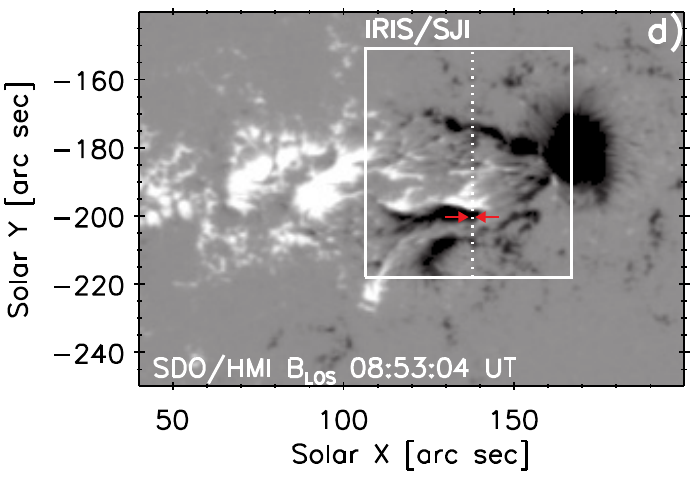 }
    \includegraphics[width=6.74cm, viewport= 0  0 334 229, clip]{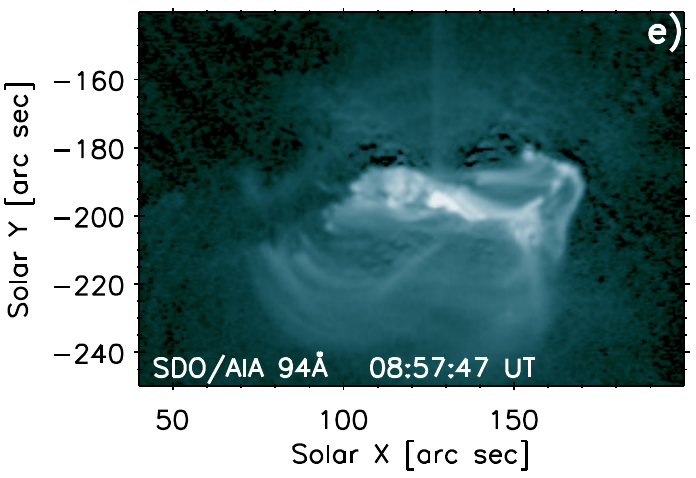 }
    \includegraphics[width=5.43cm, viewport=65  0 334 229, clip]{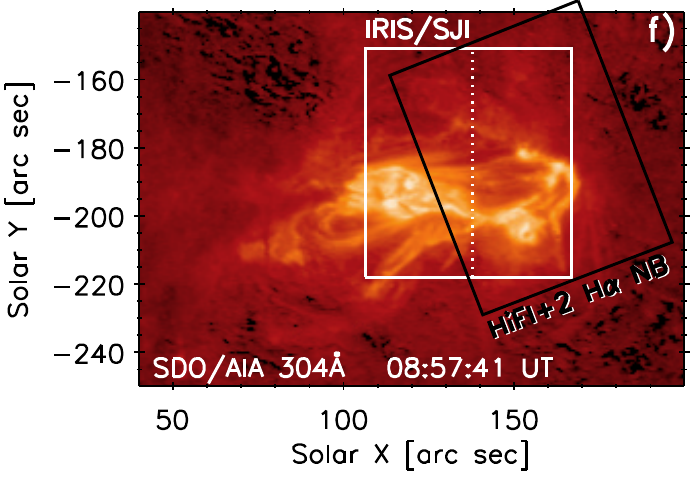 }
    \includegraphics[width=5.43cm, viewport=65  0 334 229, clip]{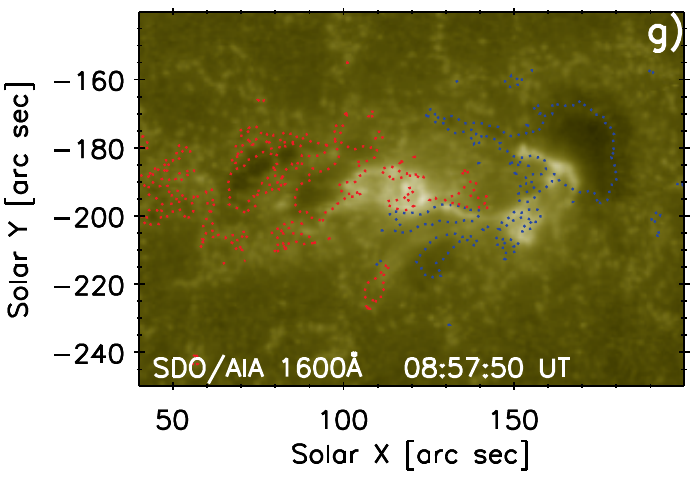 }
    \includegraphics[width=5.41cm, viewport= 0  0 240 229, clip]{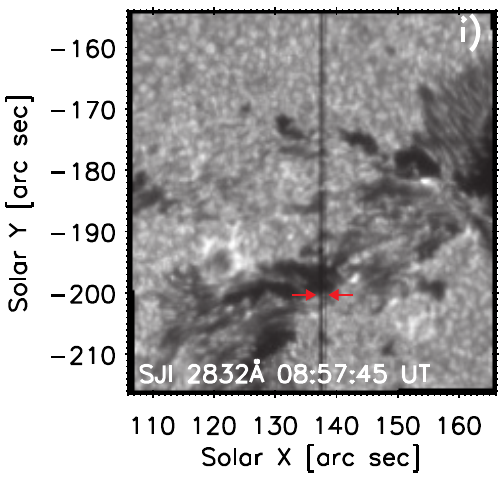 }
    \includegraphics[width=4.06cm, viewport=60  0 240 229, clip]{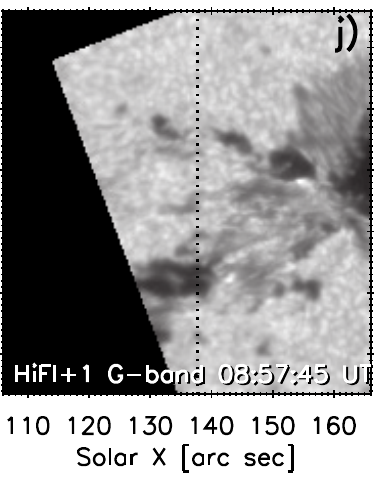 }
    \includegraphics[width=4.06cm, viewport=60  0 240 229, clip]{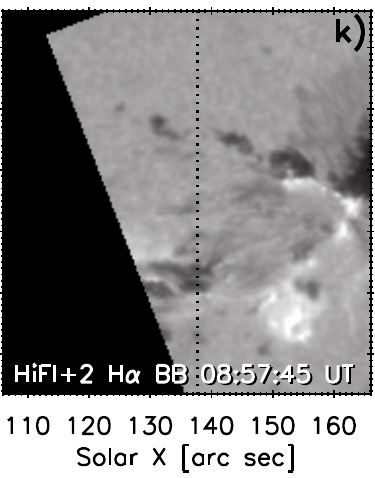 }
    \includegraphics[width=4.06cm, viewport=60  0 240 229, clip]{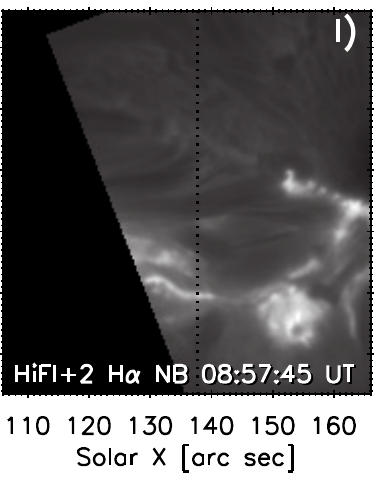 }
    \includegraphics[width=6.74cm, viewport= 0  0 334 229, clip]{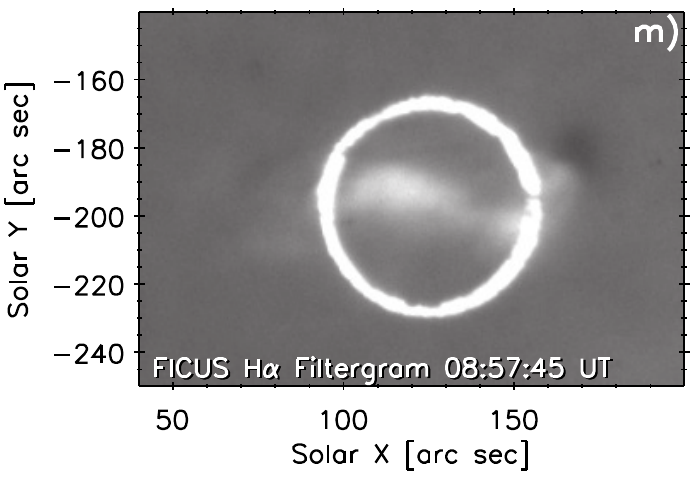 }
    \includegraphics[width=5.43cm, viewport=65  0 334 229, clip]{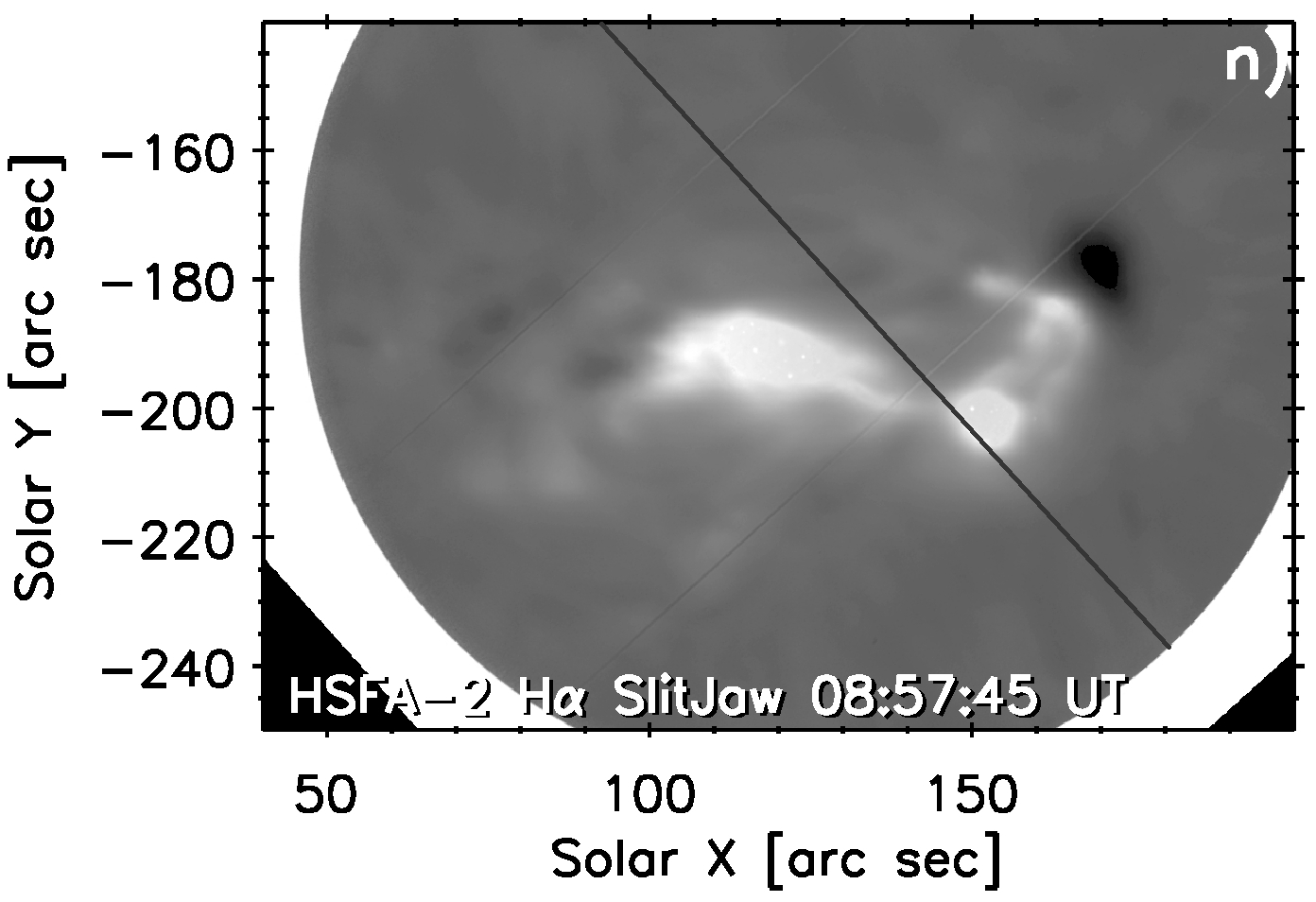}
    \includegraphics[width=5.43cm, viewport=65  0 334 229, clip]{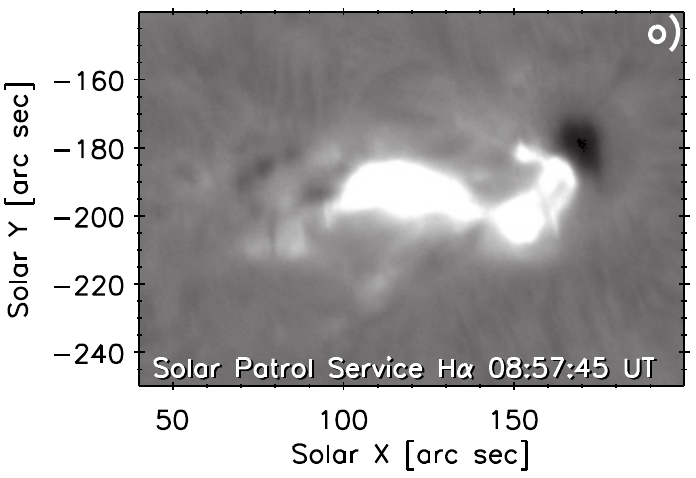 }
\caption{Overview of the 2022 May 4 M5.7--class flare as observed by multiple instruments. Panel a) shows the GOES X-ray light curve of the flare. Panels b)--g) show the context SDO/AIA and SDO/HMI observations at 08:52:45 and 08:57:45\,UT, denoted by vertical lines in panel a). Panel g) includes the contours of the polarities (positive in red and negative in blue) from panel d). The FOVs of the HiFI+2 and IRIS/SJI instruments are indicated in panel f) (only IRIS/SJI FOV in panel d), and the corresponding observations are shown in panels i)--l), with the location of the IRIS slit indicated and the HiFI+ images resampled to the IRIS resolution and FOV.
The red arrows in panels d) and i) point to the location of pixel No.\,53. Ond\v{r}ejov ground-based observations from the spectrographs FICUS and HSFA-2 and the SPS are shown in panels m)--o), respectively. The black line in panel n) indicates the HSFA-2 slit position. Animations of the AIA, IRIS, and GREGOR observations (panels e)-- l)) are available \href{https://doi.org/10.1051/0004-6361/202451219}{\textbf{online}}.}
\label{Fig:Overview}
\end{figure*}
%--------------------------

Solar flares \citep{Fletcher2011} represent the most energetic processes within the Solar System and are a consequence of solar magnetic activity. The energy released by magnetic reconnection in solar flares is converted to other forms of energy, such as the acceleration of charged particles, bulk plasma heating, plasma motions that include coronal mass ejections, amongst others. Much of the released magnetic energy is deposited by beams of accelerated particles into the solar chromosphere, which is heated and evaporates \citep[e.g.,][]{Young2013,Graham2015,Polito2016}, leading to the creation of hot (several tens of million Kelvin) flare loops higher in the atmosphere and bright flare ribbons at the chromospheric footpoints of the flare loops. These structures containing heated plasma then emit radiation in many portions of the electromagnetic spectrum \citep[e.g.,][]{Fletcher2011}. Since the energy deposit into the solar chromosphere by necessity precedes the creation of the hot flare loops, observations of the heating of the chromosphere are indispensable for understanding the energy release in solar flares. 

Due to its relatively low temperatures, around 10$^{4}$\,K, the chromosphere contains large amounts of neutral hydrogen \citep[e.g.,][]{Anzer05,Leenaarts07}. The heating of the chromosphere can produce multiple well-known spectral lines as well as continuum emission. Although line emission in chromospheric or transition region lines is prominent during solar flares and can be enhanced by orders of magnitude, continuum emission also contributes significantly to the amount of energy radiated away (see, e.g., indications from observations \citep{kretzschmar2011} and modelling \citep{Allred2005}).

Several mechanisms have been proposed for the enhancement of the continuum during solar flares, namely hydrogen bound-free and free-free transitions, Thomson scattering, and H$^{-}$ emission \citep[e.g.,][]{Heinze2014, Heinzel2017}. These mechanisms emit in the wavelength range from near ultraviolet (NUV) to near infrared (IR). Each of them may dominate in different atmospheric layers, from the photosphere to the chromosphere.  

These mechanisms require not only an increase of temperature in the emitting layers but also an increase of electron density, yet which processes increase them is still not fully understood. Some of the proposed processes are electron and/or proton bombardment, X-ray and extreme ultraviolet heating, and Alfv\'{e}n wave dissipation \citep[for discussion see, e.g.][]{Metcalf1990}. Many continuum enhancements detected during flares show good temporal and spatial correlation with hard X-ray (HXR) footpoint emission \citep[see, e.g.,][]{Neidig1993,Fletcher2007,Martinez2012, Xu2012,Milligan2014, Kleint_et_all_2016,Kuhar2016}. This suggests that the accelerated electrons that generate HXR emission and deposit their energy in the flare footpoints are also responsible for the continuum emission. However, this mechanism is probably not sufficient to explain the observed emission at $1.56\,\mu$m \citep{Xu2006}, which is considered to form at the opacity minimum below $\tau_{500}=1$ in an undisturbed atmosphere ($\tau_{500}$ is the optical depth at 500~nm). The required energy of accelerated particles that reach and affect such deep layers is typically higher than the energy determined from their HXR emission \citep[e.g.,][]{Fletcher2011}. Additionally, radiative transfer computations incorporated in flare modelling have also shown that the photosphere and the chromosphere can be radiatively coupled via photospheric heating by H$^{-}$ absorption of the hydrogen Balmer continuum, which originates in the chromosphere. This back-warming can then lead to increased photospheric H$^{-}$ radiation \citep[e.g.,][]{Machado1989}.

Dedicated modelling of solar flare radiation is required to disentangle the various mechanisms affecting continuum emission. \citet{Mauas1990} combined non-LTE (outside of Local Thermodynamic Equilibrium) modelling  and observations of several photospheric and chromospheric lines together with the visible-light continuum and constructed a semi-empirical model of a solar flare atmosphere. The formation of UV, visible, and IR continua due to electron beam heating has been modelled to various degrees of agreement with observations, for example, by \citet{Ding1999, Ding2003}, \citet{Cheng2010}, and \citet{Simoes2017}.

Studies of the continuum during solar flares also need dedicated observations and data analyses. Continuum enhancements in the UV range have been reported from space-borne observations. \citet{Heinze2014}, \citet{Joshi2021}, and \citet{Kleint_et_all_2016} identified NUV wavelength ranges observed by IRIS where continuum flare emission could be identified. These authors attributed the observed enhancements to hydrogen Balmer recombination. Similarly, \citet{Dominique2018} adopted the same assumption and analysed continuum UV flare emission detected by the PROBA-2 mission. Continuum emission and its enhancement in the optical range during strong and even weak solar flares has also been detected in filtergrams and spectral data (see, e.g., \citet{Jess2008,Kerr2014,Potts2010ApJ,Yurchyshyn2017}). Furthermore, \citet{Heinzel2017} interpreted the off-limb flare emission in the HMI/SDO pseudocontinuum as being dominated by the hydrogen Paschen recombination. \citet{Jurcak2018_flare} derived optical continuum enhancement from SOT/Hinode spectropolarimetric observations during a solar flare and proposed that it was caused by chromospheric emission.

In this work, we present an initial analysis from a unique dataset obtained during a coordinated campaign between the GREGOR telescope (campaign supported by SOLARNET), IRIS, and Hinode (HOP\,422) and instruments located at the Ond\v{r}ejov Observatory. For the first time, IRIS and GREGOR co-observed the preflare, impulsive, and gradual phase of a solar flare. The current analysis is inspired by previous studies focused on UV and optical flare continuum enhancements. Our data are unique in their UV wavelength coverage and temporal resolution, which is provided by the full read-out of IRIS NUV high-resolution spectra with a time resolution of 3\,s. Sub-arcsecond co-spatial observations in several optical wavelengths obtained with the High-resolution Fast Imager (HiFI+) at the GREGOR telescope complement the dataset. 

In Sect. \ref{Sect:data}, we describe the investigated data and present multi-wavelength light curves of the flaring atmosphere in Sect. \ref{sect:lightcurves}. The flare continuum enhancements in the optical and UV are defined in Sect. \ref{Sect:Continua}. In Sect. \ref{Sect:Temperature}, we determine the lower limit for the temperature in the flaring chromosphere above the edge of a pore by assuming that the continuum enhancement is mainly due to hydrogen recombination. Finally, the results are discussed and summarised in Sect. \ref{Sect:Discussion} and \ref{Sect:Summary}.

%________________________________________________________________
\section{Data}
\label{Sect:data}

\subsection{The 2022 May 4 flare}

On 2022 May 4, a confined M5.7-class flare occurred in a bipolar active region (AR) NOAA 13004 (Hale classification $\beta/\beta$), starting at about 08:45\,UT (Fig.~\ref{Fig:Overview}a). 
This AR was quite active, producing four other C-class flares before the M-class one, including a C1.4 starting at 07:14\,UT and C3.5 at 08:08\,UT.

The M-class flare was well observed using multiple ground-based and space-borne instruments (Fig.~\ref{Fig:Overview}). The ground-based instruments focused on the visible and infrared portion of the spectrum and involved the GREGOR solar telescope \citep{Schmidt12} located in Tenerife (Spain) as well as the spectrographs HSFA-2 \citep{Zapior_et_al_22}, FICUS \citep{Kotrc_et_al_2016}, and the Solar Patrol Service (SPS) located at the Astronomical Institute of the Czech Academy of Sciences (Ond\v{r}ejov, Czech Republic). The space-borne instruments included the Atmospheric Imaging Assembly \citep[AIA;][]{Lemen12} and Helioseismic and Magnetic Imager \citep[HMI;][]{Scherrer12} on board the Solar Dynamics Observatory \citep[SDO;][]{Pesnell12} as well as the Interface Region Imaging Spectrograph \citep[IRIS;][]{DePontieu14}. Additionally, the GOES-16 satellite and Konus spectrometer \citep{Aptekaretal1995} on board the Wind spacecraft \citep{1995_wind} provided X-ray context information.   

\subsection{Context observations from SDO/AIA} \label{Sect:AIA}

The AIA observes the Sun in 10 EUV and UV filters, spanning a large range of temperatures of the transition region, corona, and flares \citep{ODwyer10,Lemen12,DelZanna13}. We investigated the EUV (UV) data with a full cadence of 12\,s (24\,s) during the interval 07:00--11:00\,UT, covering the entire flare as well as a few additional hours for context. The AIA data have a pixel scale of 0.6\arcsec pixel$^{-1}$ and a spatial resolution of 1.5\arcsec pixel$^{-1}$ \citep{Lemen12}. The data were calibrated to level 1.5 using the \texttt{aia\_prep} routine in the IDL SolarSoft System \citep[SSWIDL;][]{FreelandHandy1998}. The stray light was deconvolved using the routine of \citet{Poduval13}, and the data were corrected for differential rotation of the Sun.

To study the flare evolution, we used the 94\,\AA~passband, which under flare conditions is dominated by the emission of \ion{Fe}{XVIII} because the hotter 131\,\AA~passband (dominated by \ion{Fe}{XXI}) is at times saturated. The 94\,\AA~indicates that the flare is a confined flare \citep[denominated Type I by ][]{Li19,Duan22}. The flare emission consists of multiple hot flare loops (Fig.~\ref{Fig:Overview}b and \ref{Fig:Overview}e). % that exhibit slipping motion \citep[see][and references therein]{Janvier13,Dudik14,Dudik16,Li14,Li15,Li16,Sobotka16,Lorincik19}. 
The system of hot flare loops grows in size during the period from about 08:47 UT to 09:10\,UT. A peak of the X-ray flare as measured by GOES is reached at 08:58\,UT (Fig.~\ref{Fig:Overview}a). 

The corresponding ribbons were captured by the 304\,\AA~and 1600\,\AA~passbands of AIA (Fig.~\ref{Fig:Overview}c, \ref{Fig:Overview}f, and \ref{Fig:Overview}g). These passbands are dominated by transition region emission of \ion{He}{II} at 303.8\,\AA\ and a \ion{C}{IV} 1550\,\AA~doublet with many other chromospheric lines producing a flare signal in the 1600\,\AA~passband as well as an about 20\% contribution of photospheric continua \citep{Simoes19}. The 1700\,\AA~passband is dominated by a multitude of chromospheric lines, with some contributions of \ion{He}{II} \citep[see Fig. 6 of][]{Simoes19}. Finally, the 94\,\AA~channel of AIA, which we used to identify the flare loops (Fig.~\ref{Fig:Overview}b and \ref{Fig:Overview}e), is dominated by the \ion{Fe}{XVIII} \citep{ODwyer10}. As there are several systems of hot flaring loops, the structure of the ribbons is quite complex and has some small-scale substructure.%, which is typical of Type I confined flares \citep[cf.,][]{Li19}.  

Since AIA is a full-disc instrument with good pointing, we used it to determine the absolute solar coordinates of the datasets with the following procedure. First, the 171\,\AA~channel of AIA was taken as a reference, and other AIA passbands were aligned to it. This step corrected for minor spatial shifts in passbands, such as 304\,\AA. The 1600\,\AA~and 1700\,\AA~passbands, which dominantly show the solar photosphere and not the transition region nor corona, were aligned to 304\,\AA~using the location of flare ribbons. The IRIS/SJI data (see Sect. \ref{Sect:IRIS}) were aligned to AIA 1600\,\AA~using the location of photospheric features such as pores and sunspots. Similarly, the H$\alpha$ data from the Ond\v{r}ejov observatory (Sect. \ref{Sect:Ondrejov}) were also aligned to 304\,\AA~and 1600\,\AA~data through the use of the locations of ribbons, sunspots, and pores. In this manner, a co-aligned set of data from multiple instruments was obtained, with uncertainties below $0.6\arcsec$.

% Since AIA is a full-disc instrument with good pointing, we determined the absolute solar coordinates using the following procedure. First, the 171\,\AA~channel of AIA is taken as reference, and other AIA passbands are aligned to it. This corrects for minor spatial shifts in passbands such as 304\,\AA. The 1600\,\AA~and 1700\,\AA~passbands, which dominantly show the solar photosphere and not transition region or corona, are aligned to 304\,\AA~using the location of flare ribbons. The IRIS/SJI and SDO/HMI (see Sects. \ref{Sect:IRIS} and \ref{Sect:HMI}, respectively) data are aligned to AIA 1600\,\AA~using the location of photospheric features such as pores and sunspots. Similarly, the H$\alpha$ data from Ond\v{r}ejov observatory (Sect. \ref{Sect:Ondrejov}) are also aligned to 304\,\AA~and 1600\,\AA~data, using the locations of ribbons, sunspots, and pores. In this manner, a co-aligned set of data from multiple instruments is obtained, with uncertainties below 0$\farcs$6. 

%
\subsection{IRIS observations}
\label{Sect:IRIS}

On 2022 May 4, IRIS observed in sit-and-stare mode between 07:30 and 13:00\,UT as part of the Hinode Operation Plan (HOP) 422 (OBSID ref. 3884855852). During the whole period of observations, 5984 exposures were taken, with an approximate cadence of 3\,s. In order to achieve a good temporal resolution, we had to sacrifice spatial resolution in all datasets ($\times$2 binning) and spectral resolution. IRIS provided high-resolution spectra in the NUV channel ($\times$2 binning) and low-resolution spectra in the far-ultraviolet (FUV) channel ($\times$8 binning) for all exposures. In this work, we do not analyse the FUV spectra due to the low resolution as well as the abundance of lines that hindered the study of the continuum.

The IRIS NUV channel spans from 2783.63 to 2835.11\,\AA, and it includes Mg~II lines, their far wings, and the continuum at the longer wavelength end of the spectral window. In the full read-out, the spectra contain 1012 wavelength points, with a spectral sampling of 0.051\,{\AA}. We converted the signal given in data number (DN) units to absolute energetic units (erg\,s$^{-1}$\,cm$^{-2}$\,sr$^{-1}$\,\AA$^{-1}$) in all NUV spectra using the IRIS radiometric calibration (see Appendix \ref{Sect:Appendix_iris_calib}). The spatial coverage of the slit corresponds to the Y-size of the slit-jaw images (SJIs; 62\arcsec, 203 pixels) and gives a spatial sampling of around 0.3\arcsec pixel$^{-1}$. The SJIs are only available in the 2832\,{\AA} channel, with a field of view (FOV) of 60\arcsec\,$\times$\,62\arcsec and a pixel scale of 0.3\arcsec. An example is presented in Fig.~\ref{Fig:Overview}i. 

An animation of the SJI 2832\,{\AA} frames (available \href{https://doi.org/10.1051/0004-6361/202451219}{\textbf{online}}) shows a rather complex evolution and motion of pores and small sunspots, while the bigger leading sunspot kept a relatively stable position at the western edge of the SJI FOV. The ribbons started to be visible at 08:54\,UT for a few minutes and were located above the umbrae and penumbrae of the sunspots.

The transmission function of the 2832\,{\AA} filter peaks in the longer wavelength end of the NUV spectral range and covers mainly the far wings of Mg II lines, where the continuum dominates. However, many other weak lines, as well as the Mg II k line, can contribute significantly to the total signal in the SJI 2832\,\AA~passband during flares \citep{Kleint2017}. Therefore, we considered the flare ribbons observed in this passband as regions in which continuum emission might be dominant but not the only contributor.

\subsection{GREGOR observations}
\label{Sect:GREGOR}

GREGOR observed the flaring region from 08:45 to 09:16\,UT. The GREGOR Infrared Spectrograph \citep[GRIS;][]{Collados12} started scanning the AR at 08:45\,UT, but from 08:50 to 08:52\,UT, we re-adjusted the FOV and changed the observing mode. This was a period of a mediocre seeing (Fried parameter $r_0 = 5$--10\,cm), but thanks to the GREGOR adaptive optics system \citep{Berkefeld12}, the pointing was stable and the image quality acceptable. Both the spectra and images of the flare were acquired.

The spectra were retrieved with a slit in sit-and-stare mode with GRIS set to the 1.08 $\mu$m wavelength band, which includes photospheric and chromospheric lines. Since the GRIS slit was not co-spatial with the IRIS slit at any point, the GRIS spectral observations are not analysed here and are left for future work. 

Images of the flare were recorded by the improved High-resolution Fast Imager \citep[HiFI+;][]{Denker23}, which is composed of three sets of two synchronised cameras. HiFI+1 contains two sCMOS cameras with an FOV of $70.7\arcsec \times 59.6\arcsec$ and a pixel scale of $0.028\arcsec\mathrm{pixel}^{-1}$, plus interference filters with wavelength bands at the G-band 4307~{\AA} and blue continuum 4506~{\AA} and passbands of 11~{\AA}. HiFI+2 has two Imager M-lite 2M CMOS cameras with an FOV of $76.5\arcsec \times 60.5\arcsec$ and a pixel scale of $0.050\arcsec\mathrm{pixel}^{-1}$, plus a broadband interference filter and a narrowband Lyot filter with wavelength bands at H$\alpha$ 6563~{\AA} and passbands of 7.5 and 0.6~{\AA}, respectively. HiFI+3 has two Imager M-lite 2M CMOS cameras with an FOV of $48.2\arcsec \times 30.8\arcsec$ and $76.5\arcsec \times 60.5\arcsec$ and pixel scales of $0.025\arcsec\mathrm{pixel}^{-1}$ and $0.050\arcsec\mathrm{pixel}^{-1}$, respectively, plus interference filters with wavelength bands at \ion{Ca}{II}~H 3968~{\AA} and the TiO molecular band 7058~{\AA} and passbands of 9 and 11~{\AA}, respectively. Examples of the HiFI+ observations in the G-band and the H$\alpha$ broad- and narrowbands are shown in Fig.~\ref{Fig:Overview}j, \ref{Fig:Overview}k, and \ref{Fig:Overview}l. 

The CMOS and sCMOS cameras allow for a fast image-acquisition rate: HiFI+1 recorded at 49~Hz, and HiFI+2 and HiFI+3 recorded at 100~Hz. The exposure times were constant for HiFI+1 (G-band: 2~ms; blue continuum: 1~ms) and HiFI+3 (TiO: 0.5~ms). In order to avoid saturation, the exposure times were reduced from 9~ms to 7~ms at 08:56~UT for HiFI+2 (broad- and narrowband) and from 8~ms to 6~ms at 08:52~UT for HiFI+3 (Ca~II~H). Even though each of the HiFI+ camera sets is controlled by a different computer temporally synchronised with a time server, we found a temporal delay of $\sim 6.5$~s in the HiFI+3 images with respect to HiFI+1 and HiFI+2. The analysis was done correcting for this delay.

The images were calibrated using the \texttt{sTools} image processing pipeline \citep{2017_stools}\footnote{\url{gitlab.aip.de/cdenker/stools}} for the full extent of the HiFI+ acquisition rate capabilities. In other words, no frame selection was done.

In order to carry out a co-spatial analysis of GREGOR and IRIS observations, HiFI+ images were aligned to the IRIS SJI frame closest in time using routines in the SSWIDL. First, HiFI+ images were smoothed with a Gaussian filter to mimic IRIS spatial resolution. Preliminary linear transformation parameters ($P_0$, $Q_0$; rotation, scale, and shift per coordinate) were retrieved by \texttt{caltrans.pro}, which used as an input reference points created with \texttt{setpts.pro}. Next, we masked the pixels with a high signal caused by the flare (bright pixels) and the pixels where the slit was located in IRIS SJIs. A final alignment was done by the SSW routine \texttt{auto\_align\_images.pro} in two runs. The initial run imported $P_0$, $Q_0$ and calculated the $P_1$, $Q_1$ that produced the minimum cross correlation between IRIS and HiFI+ images using the faster amoeba minimisation algorithm. The second run corrected HiFI+ images for warping and employed the more robust Powell minimisation algorithm \citep{Pressetal1992}. In this way, we obtained the aligned HiFI+ images together with the $P_f$, $Q_f$ (final) transformation parameters.

Each HiFI+ image was aligned independently, except for the entirely chromospheric HiFI+2 narrowband images. These were transformed and rescaled to the IRIS FOV by \texttt{poly\_2d.pro}, which applied the $P_f$, $Q_f$ parameters obtained from the synchronous HiFI+2 broadband images. As seen in Fig.~\ref{Fig:Overview}, the HiFI+ images do not completely overlap with IRIS SJIs; however, the IRIS slit is completely covered by almost all of the HiFI+ imagers, except for HiFI+3 (\ion{Ca}{II}~H).

\subsection{SDO/HMI observations}
\label{Sect:HMI}

The SDO/HMI synoptic observations were used to estimate the excess in the optical continuum; therefore, we required a dataset that allowed us to study the intensity in the continuum. The standard product capturing the intensity of the pseudo-continuum around the \ion{Fe}{I} 6173\,\AA{} line is the $I_{\rm c}$ product available via JSOC\footnote{\url{http://jsoc.stanford.edu}} with a cadence of 45~s. This product results from the reconstruction of the spectral line from a non-trivial sequence of line scans in various light-polarisation states, and it is known not to represent the true continuum signal very well in solar flares \citep{Svanda_et_at_2018}.

To overcome this issue, we used level\,1 data, the filtergrams, which are available upon request from JSOC. The HMI scans the iron line at six nominal positions. The filtergrams were aligned to IRIS/SJI using the \texttt{reproject\_to} routine from the SunPy Project \citep{SunPy}. In the following, we focus only on the outer positions, that is, those representing $-172$~m\AA{} and $+172$~m\AA{} shifts from the nominal spectral-line position. A continuum enhancement should be registered at all scan positions; therefore, utilising the measurements far in the spectral-line wings allowed us to separate the continuum enhancement from a possible brightening in the spectral line. Unfortunately, these filtergrams were not recorded simultaneously, and their time stamps differ due to the sequence scan performed by HMI. To have a pseudo-continuous series, we interpolated both filtergram series to a uniform series with a cadence of 30~s. Then, we averaged the blue-most and the red-most filtergrams to represent the evolution of the continuum. 

%Finally, HMI data were aligned to IRIS/SJI using the \texttt{reproject\_to.py} routine from the SunPy Project \citep{SunPy}.   

\subsection{\texorpdfstring{H$\alpha$}{TEXT} and spectral observations from Ond\v{r}ejov}
\label{Sect:Ondrejov}

The flare was also observed by ground-based instruments located at the Ond\v{r}ejov observatory. These include an H$\alpha$ telescope and two spectrographs that operate in the visible and infrared portions of the solar spectrum. These instruments supported the coordinated campaign of GREGOR and HOP~422 by observing the same target regions.

The Solar Patrol Service (SPS) observes the Sun in the photospheric continuum and H$\alpha$ line filters, both in two modes: full-disc and detail of a selected AR. On 2022 May 04, SPS observed the AR with an H$\alpha$ filter using a 210mm f/16.3 achromatic refractor, producing images with a 0.44$\arcsec$ plate scale and 1\,s temporal cadence. As the observations are not in a final configuration and not yet flare-optimised, the flare ribbons in the H$\alpha$ SPS image can become saturated (see Fig.~\ref{Fig:Overview}o).

The Horizontal-Sonnen-Forschungs-Anlage-2 (HSFA-2; horizontal device for solar research) is a high-dispersion spectrograph. It simultaneously observes five spectral regions in high spectral resolution along with slit-jaw H$\alpha$ images of the target region. The last upgrade was done in 2020. In the current setup, it observes with an FOV with a radius of $\sim$280$\arcsec$. An example of an SJI taken during the campaign is shown in Fig.~\ref{Fig:Overview}n. The grey line indicates the slit position. More details about the spectrograph can be found in \citet{Zapior_et_al_22} and \citet{kotrc_09}.

The Flare Intensity Continuum Ultrawide Spectrograph (FICUS) is a new low-dispersion spectrograph based on a setup described by \citet{Kotrc_et_al_2016}. The instrument has been relocated so that it is now fed by its own dedicated coelostat. The FICUS instrument provides spectra in the 3500-9300\,\AA~wavelength range and H$\alpha$ images of a selected region with a time resolution on a (sub)second timescale and a plate scale of about 0.5$\arcsec$. The FICUS design employs an image selector and several circular diaphragms of varying aperture to select the observed area. An example of a context H$\alpha$ image from FICUS is shown in Fig.~\ref{Fig:Overview}m. There, the area observed by the spectrograph is delimited by the bright white circular rim of the diaphragm, denoting the selected $61\arcsec$ aperture diameter.

The coalignment of the three H$\alpha$ observations from Ond\v{r}ejov was done manually, first the SPS with AIA 1600\,\AA~and then FICUS and HSFA-2 with respect to SPS. The SPS data were of maximum importance during the GREGOR campaign, where observers benefited from the real-time SPS H$\alpha$ full-disc and detailed observations from which to choose an interesting target for GREGOR. Due to the not-yet optimised configuration of SPS, its data were not used for further analysis. Spectral data from HSFA-2 were not analysed in this work either since the HSFA-2 slit did not cross the flare region captured by the IRIS slit. Preliminary analyses of the FICUS spectra did not show any continuum flare enhancement, probably due to the small portion of the flare area within the FICUS FOV. 

\subsection{X-ray observations}

During the impulsive phase, the flare emission in the X-ray domain was only observed by the GOES-16 satellite and the Konus-Wind instrument. The Wind spacecraft orbits in interplanetary space (L1 Lagrange point), which provides a stable background for Konus omnidirectional observations. Konus detected hard X-ray emission in the 19~-~80~keV energy range (G1 channel) during the so-called waiting mode. Therefore, the data are available as a count rate light curve in the wide energy channel G1 with an accumulation time of 2.944~s \citep{Lysenkoetal2022}.

%%%%%%%%%%%%%%%%%%%%%%%%%%%%%%%%%%
%%%%%%%%%%%%%%%%%%%%%%%%%%%%%%%%%%

\section{Multi-wavelength evolution of a flare kernel}
\label{sect:lightcurves}

\begin{figure*}[h!]
    \centering 
    \includegraphics[width=\textwidth]{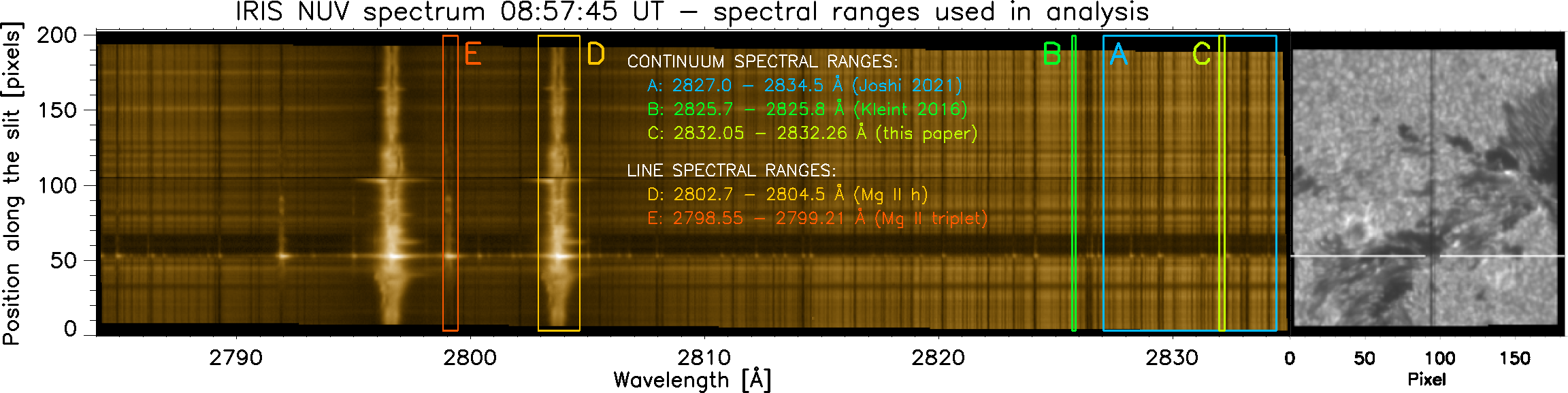}
    \caption{IRIS NUV spectrum recorded at the time of maximum continuum emission (08:57:45\,UT) along the full IRIS slit with colour boxes marking the spectral ranges (A, B, C, D, and E) used for the light curve plots and for the continuum analysis. On the right is a synchronous IRIS SJI snapshot with a horizontal white line marking the height of pixel No.\,53. Animations of the IRIS data are publicly available 
    \href{https://www.lmsal.com/hek/hcr?cmd=view-event&event-id=ivo\%3A\%2F\%2Fsot.lmsal.com\%2FVOEvent\%23VOEvent_IRIS_20220504_073000_3884855852_2022-05-04T07\%3A30\%3A002022-05-04T07\%3A30\%3A00.xml}{here}. } 
\label{fig:iris_sp_boxes}
\end{figure*}

Our aim is to study the flare continuum enhancement by exploiting the high temporal cadence of HiFI+ and IRIS datasets and the full read-out of the NUV spectra provided by IRIS. The IRIS slit was positioned above the polarity inversion line of the AR, and it crossed quiet-Sun regions, pores, and penumbral filaments. Gentle brightenings in the shape of flare ribbons were visible between 08:54 and 09:01\,UT. The brightest parts of the flare ribbons were located out of the IRIS slit position (see Fig.~\ref{Fig:Overview}i). Nonetheless, there was a faint flare emission crossing the slit at around 08:57:30\,UT. This emission occurred on the border of a pore, and it was located in the pixel No.\,53 (counting from the bottom) along the slit. The position of pixel No.\,53 is marked with red arrows in Fig.~\ref{Fig:Overview}i. We investigated the multi-wavelength temporal evolution of this pixel in order to extract the flare continuum emission.

\subsection{IRIS near-ultraviolet bands}
\label{sect:IRIS_bands}

The maximum flare continuum enhancement on the IRIS slit was reached at 08:57:45\,UT. The corresponding spectra along the IRIS slit are plotted in Fig.~\ref{fig:iris_sp_boxes}. One can see a spatially narrow enhanced emission at the position Y~=~53 in the whole spectral window. This enhancement is manifested by a strong emission in \ion{Mg}{II}~h, k, and triplet lines as well as by a gentle increase of the continuous emission in the whole spectral range. Moreover, brightenings are also present in many other narrow lines. Pixel No.\,53 is located at the edge of a dark structure that corresponds to a pore of the size $\sim$3.3\arcsec that spans from pixels No.\,55 to 65.

The location of pixel No.\,53 at the edge of the pore introduces complications in studying the long-term evolution of the flare continuum. This is because the IRIS pointing wobbles by about $0.6\arcsec$ in the solar $Y$ direction during the HOP\,422 (see Appendix \ref{Sect:Appendix_wobble}). Fortunately, the flare occurred in one of the troughs of the wobble, when the pointing changes were very slow between about 08:40 and 09:00\,UT (Fig.~\ref{Fig:Appendix_IRIS_pointing}). In the following, we restrict our analysis of the IRIS flare spectra to this time interval.

\begin{figure*}
    \centering
    \includegraphics[width=1.0\textwidth,viewport= 0 0 566 225, clip]{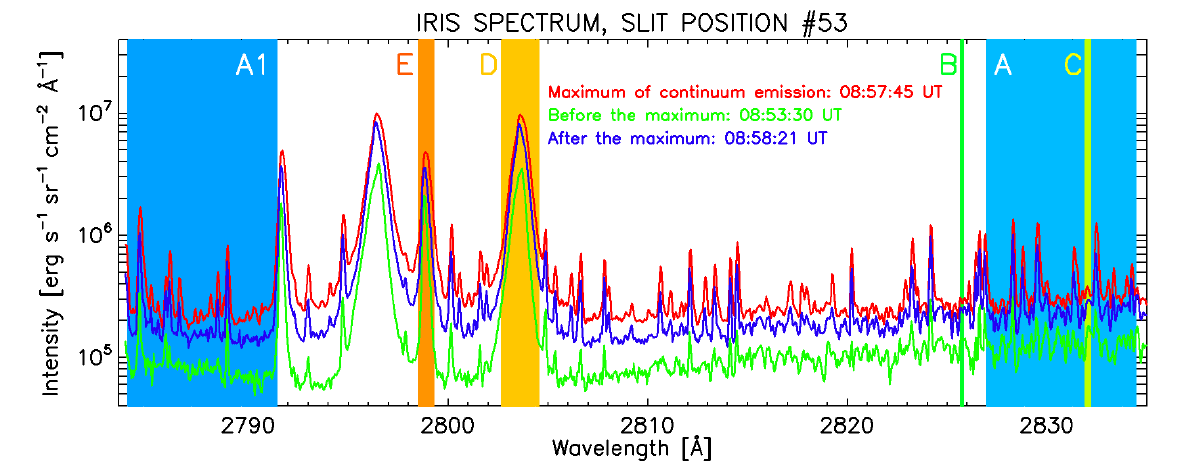 }
    \caption{IRIS NUV spectral profiles obtained at the position of the flare kernel (pixel No.\,53 along the IRIS slit). The three curves represent the spectrum at the time of maximum enhancement of the continuum emission at 08:57:45\,UT (red curve) and at times before and after the maximum (green and blue curves respectively). With different colours (see description in Fig.~\ref{fig:iris_sp_boxes}), we mark the spectral ranges A - E. We also marked the A1 range used in \citet{Joshi2021} but excluded from our analysis (see the text for details). } 
    \label{fig:iris_sp_pre_post_v2}
\end{figure*}

\begin{figure*}
    \centering
    \includegraphics[width=1.0\textwidth,viewport= 0 0 566 225, clip]{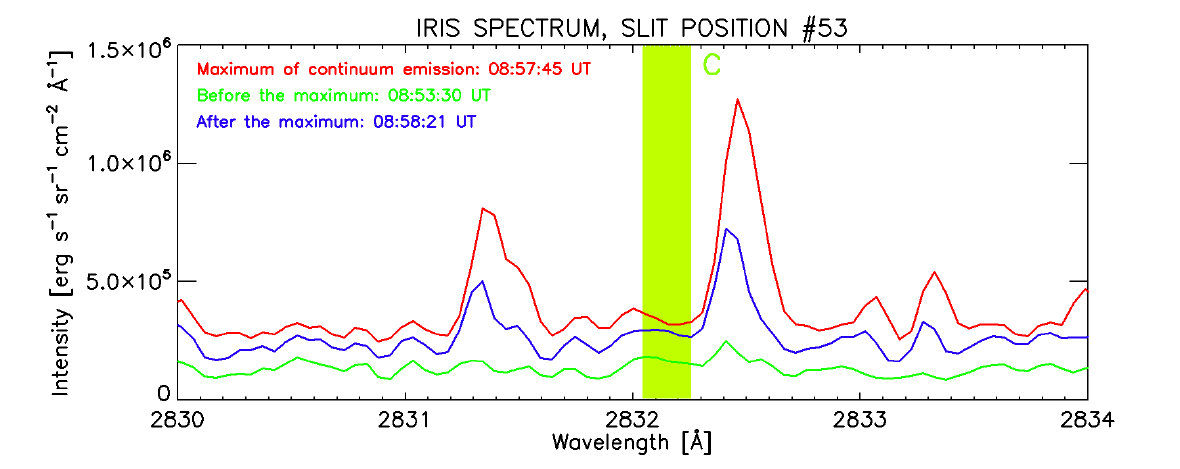 }
    \caption{Zoom-in of Fig.~\ref{fig:iris_sp_pre_post_v2} to show the spectral range used in this paper for the continuum enhancement analysis, IRIS band C (green box).  }
    \label{fig:iris_sp_pre_post_zoom}
\end{figure*}

For a detailed analysis of the temporal evolution of the flare emission within the NUV spectral window, we delimited the spectral bands A, B, C, D and E, inside which the emission was determined by averaging the intensity within the band. The positions of these bands in the spectra are presented in Fig.~\ref{fig:iris_sp_boxes}. Bands A, B, and C are meant to represent continuum emission, and bands D and E represent line-dominated emission. Band A was defined by \citet{Joshi2021} and covers the NUV spectrum between 2827.0 and 2834.5\,{\AA}, far from the \ion{Mg}{II} lines. This band contains many other narrow lines, which exhibited enhanced emission in the flare kernel. Band B is much narrower (2825.7 -- 2825.8\,{\AA}) and was defined by \citet{Heinze2014}. It is located in a spectral region where no absorption lines are present. Finally, we defined band C as spanning the range 2832.05 -- 2832.26\,{\AA}. This band does not contain any spectral line and is located far away from \ion{Mg}{II} lines, where the contribution from their far wings is negligible. Prominent \ion{Mg}{II} lines can exhibit strong and wide wings during solar flares \citep[see, e.g.,][]{Liu2015}; therefore, a large spectral distance from them is needed. In addition, we defined band D (2802.7 -- 2804.5\,{\AA}) and band E (2798.55~--~2799.21\,{\AA}), which represent the emission in the \ion{Mg}{II}~h line and in the \ion{Mg}{II} triplet, respectively. 

In Fig.~\ref{fig:iris_sp_pre_post_v2}, we present a photometric cut for pixel No.\,53, where we plot the spectrum before, during, and after the maximum intensity enhancement. We marked the spectral bands defined above with vertical colour boxes. We clearly observed that the spectrum intensity at the flare maximum is significantly enhanced in the whole spectral range with respect to the pre- and post- flare emission. The maximum flare emission is $\sim$2.4 times stronger than the preflare intensity and $\sim$1.2 times stronger than after the maximum. 
To get a better understanding of the here defined band~C, we zoomed-in on the spectra from Fig.~\ref{fig:iris_sp_pre_post_v2}. Figure~\ref{fig:iris_sp_pre_post_zoom}
shows that band~C is not influenced by the far wings of the \ion{Mg}{II} lines nor other weaker lines. Therefore, any enhancement observed would be due to an enhancement of the continuum. 

It is worth mentioning that in \citet{Joshi2021}, the authors investigated the flare continuum that defined not only band A but also another spectral band, A1, that covered 2784.0 -- 2791.5\,{\AA} on the short-wavelength side from Mg~II lines. This A1 band is marked as a blue box in Fig.~\ref{fig:iris_sp_pre_post_v2}, where it is clear that it contains many weak lines. The enhancements of these lines in band A1 are visible as small brightenings in pixel No.\,53 in Fig.~\ref{fig:iris_sp_boxes}. Consequently, we decided not to use this spectral range for the continuum enhancement analysis. 

\subsection{Light curves}
\label{sect:Ligthcurves}

\begin{figure*}
\centering
\sidecaption
\includegraphics[width=0.7\linewidth]{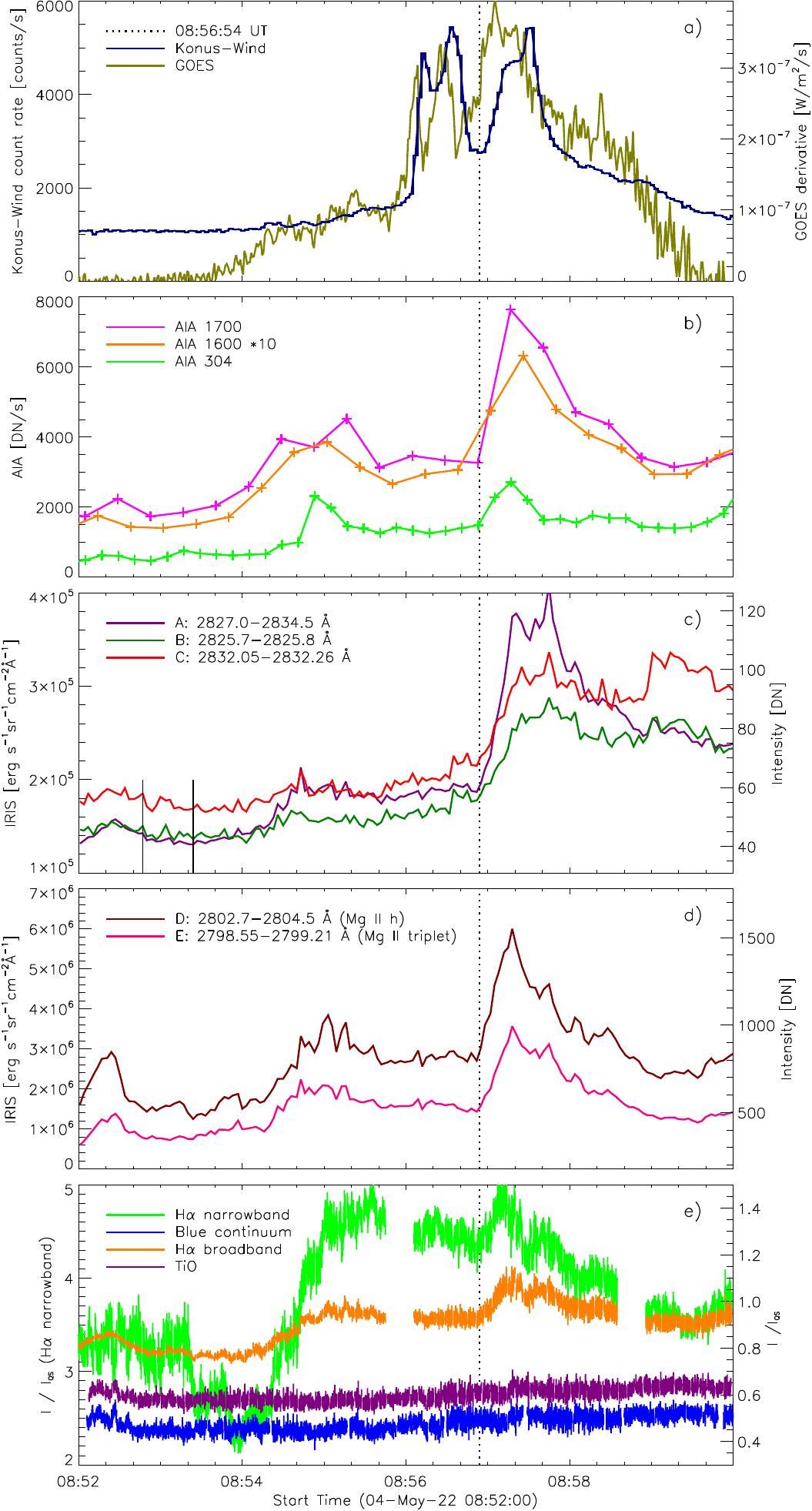 }
\caption{Light curves in several wavelength bands corresponding to spatially integrated emission (panel a) and pixels co-spatial with the pixel No.~53 along the IRIS slit (panels b - e).
Panel a: Konus count rate in the G1 channel and GOES time derivative. Panel b: Emission in several SDO/AIA filters with the 1600\,\AA\ filter data scaled by ten for plotting purposes. Panels c - d: calibrated intensity averaged in the IRIS spectral bands A, B, C (panel c; continuum emission), D, and E (panel d; line emission). Panel e: Emission relative to the quiet-Sun values in several GREGOR HiFI+ filters. The preflare time interval for evaluating the IRIS C band preflare emission is delimited by two solid vertical lines in panel c). The vertical dotted line indicates the start of the last HXR pulse.}
\label{fig:all_light_curves}
\end{figure*}

We limited the light curves to the period between 08:52 and 09:00\,UT. These times include the preflare, the impulsive phase, and part of the gradual phase. Figure~
\ref{fig:all_light_curves} summarises the multi-wavelength temporal evolution of the flare kernel located at pixel No.\,53 of the IRIS slit.

Despite a lack of spatial information in flare hard X-ray sources, we were able to identify the flare impulsive phase using spatially integrated X-ray emission. The Konus count rate in the 19~-~80 keV energy range corresponds to energies generally produced by non-thermal electrons accelerated during the flare. In addition, the time derivative of the soft X-ray emission detected by GOES was used as a proxy for the presence of non-thermal particles, assuming the Neupert effect \citep{Neupert1968}. The time variations of these light curves, shown in Fig.~\ref{fig:all_light_curves}a,
agree reasonably well and suggest that the impulsive phase and the related hard X-ray emission started at $\sim$~08:56~UT with two short spikes of several-second duration. Those spikes were followed by a third, last hard X-ray pulse starting at $\sim$~08:56:54~UT.

The AIA flare emissions in pixels co-spatial to pixel No.\,53 of the IRIS slit are depicted in Fig.~\ref{fig:all_light_curves}b. All AIA passbands (1700, 1600, and 304) show a gradual increase of intensity, with a first peak of emission at around 08:55\,UT. This was followed by a second more intense peak at around 08:57:20\,UT. The coarse temporal cadence of AIA datasets does not allow for a precise determination of the flare maximum enhancements, but they provide context.

Regarding the IRIS NUV emission, Fig. \ref{fig:all_light_curves} contains the flare evolution in all the defined bands A -- E. The continuum-dominated bands (Fig.~\ref{fig:all_light_curves}c) display a sudden increase of emission at $\sim$08:57\,UT. All the curves exhibit a first peak at 08:57:20\,UT and a second stronger peak at 08:57:45\,UT. The whole increase of the emission was observed for 80~s, starting from 08:57\,UT.

It is worth noting that bands B and C show similar temporal evolution, while curve A differs from them both. We interpret this situation in the context of the definition of the spectral bands. Band B and the here defined band C contain only continuum emission, without spectral lines. On the contrary, band A also contains many narrow spectral lines (see Fig.~\ref{fig:iris_sp_boxes}) that went into emission during the flare brightening in pixel No.\,53. It is difficult to extract pure continuum emission from such a wide spectral band. Line emission may have a different temporal evolution than continuum emission due to the different mechanisms responsible for their formation and their different formation heights. Bands B and C display an offset that is approximately constant during the whole analysed time, and it is caused by the different wavelength distance from the \ion{Mg}{II} lines. 

In Fig.~\ref{fig:all_light_curves}d, we have plotted the temporal evolution of the \ion{Mg}{II} line-dominated emission in bands D and E. These curves are very similar to each other. They show a first enhancement at around 08:55\,UT that is co-temporal to the first enhancement observed in the AIA light curves. Bands D and E also show one main maximum at 08:57:15\,UT. Unlike the continuum emission, which is formed lower in the solar atmosphere, the \ion{Mg}{II} line emission comes mainly from the chromosphere located higher and in a non-LTE regime.   

It is interesting to note that within the analysed period, there is a second intensity peak around 08:59\,UT that is only present in the continuum emission (Fig.~\ref{fig:all_light_curves}c). This secondary maximum occurs at a time when the IRIS pointing started to change due to the wobble in the $Y$-axis direction (Appendix \ref{Sect:Appendix_wobble}). This means that the secondary maximum could contain contributions from different, brighter photospheric areas as a result of the change in pointing plus contributions from the intrinsic evolution of the pore. Therefore, the secondary maximum does not necessarily reflect the evolution of the flaring continuum itself.

Light curves of optical emissions as detected by several GREGOR/HiFI+ filters corresponding to IRIS pixel No.\,53 are shown in Fig.~\ref{fig:all_light_curves}e. Both H$\alpha$ filters show an evolution similar to that in the IRIS bands D and E (line emission), namely after $\sim$\,08:54\,UT. On the other hand, the light curves of the continuum filters display intensity enhancements at $\sim$08:57:20\,UT that are co-temporal to bad seeing conditions. That is, the intensity from the dark pore was contaminated by the nearby brighter granulation. All GREGOR/HiFI+ light curves in Fig.~\ref{fig:all_light_curves}e were scaled by time-dependent quiet-Sun values (from a quiet-Sun region in the HiFI+ cameras FOV). The gap in the H$\alpha$ data at $\sim 08:56$~UT is due to a change in exposure time.

Overall, the third hard X-ray pulse ($\sim$~08:56:54~UT) is co-temporal to the IRIS continuum and line bands and to the H$\alpha$ broad- and narrowband brightenings. Moreover, brightenings are also prominent in the AIA 304, 1600, and 1700\,\AA\  light curves. Because these AIA passbands correspond to emission from transition region, chromosphere, and photosphere, we conclude that the co-temporal increase of emission in the position of IRIS pixel No.~53 with the HXR pulse was related to the heating of the solar atmosphere at that position by non-thermal electrons.

\section{Continua enhancements}
\label{Sect:Continua}

\subsection{Near-ultraviolet continuum in IRIS}
\label{Sect:IRIS_continuum}

We extracted the continuum enhancement due to the flare heating by subtracting the preflare emission from the flare emission in the here defined band C between 08:56:54 and 08:58:50\,UT. The preflare emission was calculated by averaging the intensity in a time interval indicated in Fig.~\ref{fig:all_light_curves}c. We note that X-ray and UV emissions at that time interval are also below flare values, as seen in Fig.~\ref{fig:all_light_curves}a and \ref{fig:all_light_curves}b. However, the preflare time interval precedes the start of the flare seen in the GOES X-ray emission derivative by $\sim 1$~min only.
The IRIS band C preflare value is $(1.730\pm 0.061)\times 10^{5}$~erg\,s$^{-1}$\,cm$^{-2}$\,sr$^{-1}$\,\AA$^{-1}$, determined as the mean and standard deviation of the intensities during the preflare time interval. The standard deviation was further used as an estimate of the uncertainty of the IRIS continuum intensities.

\subsection{Optical continuum }\label{Sect:optical_continuum}

The observations from GREGOR/HiFI+ and SDO/HMI were used to estimate the limit on the enhancement in the optical continuum. Both instruments recorded the intensity in arbitrary units. In order to comply with the theoretical model for the continuum, we needed to convert those values into physical units using a simplified procedure similar to that described by \citet{Kleint_et_all_2016} for HMI. The authors first determined the disc-centre intensity in the instrumental units. In order to do so, they used a small box around the disc centre from the same filtergrams they used for the construction of the continuum light curve. Then they compared this value to the atlas \citep{1994svsp.coll...37N} value of the continuum near the \ion{Fe}{I} 6173\,{\AA} line. The corresponding atlas value was $0.315\times 10^7$~erg\,s$^{-1}$\,cm$^{-2}$\,sr$^{-1}$\,\AA$^{-1}$. Assuming the linearity of the charge-coupled device (CCD) chip, the ratio of the two numbers allowed them to convert HMI measurements into the physical units using this simplistic approach. 

\begin{figure}
    \centering
    \includegraphics[width=0.49\textwidth, trim={0 1.6cm 0 0}, clip]{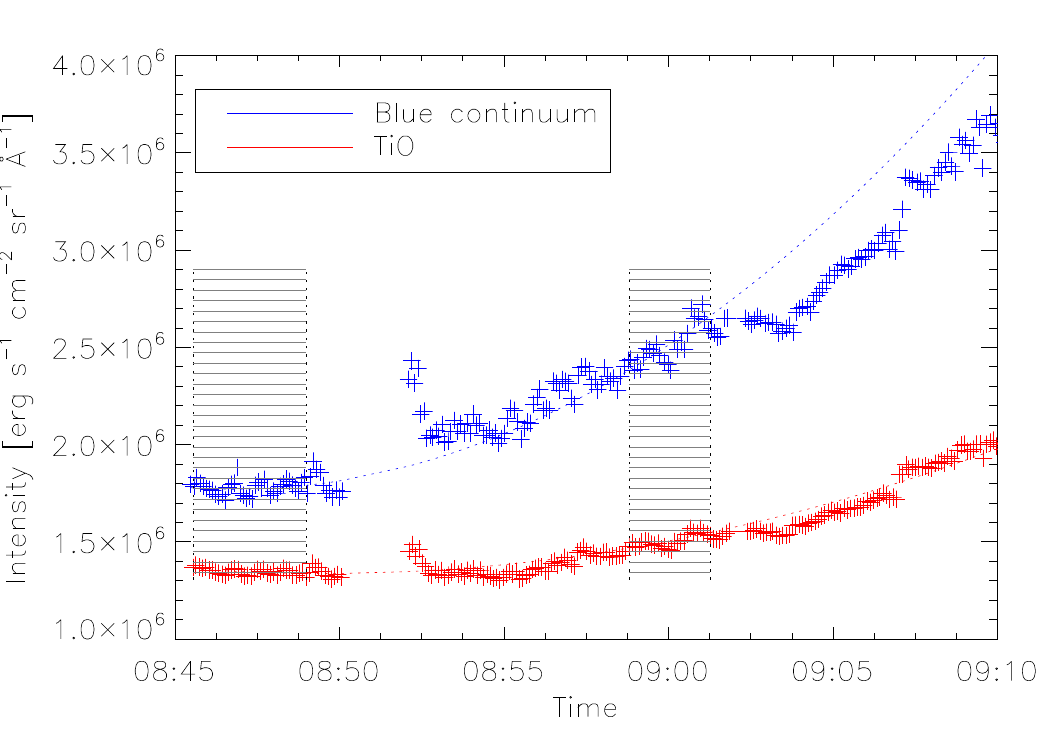 }
    \includegraphics[width=0.49\textwidth]{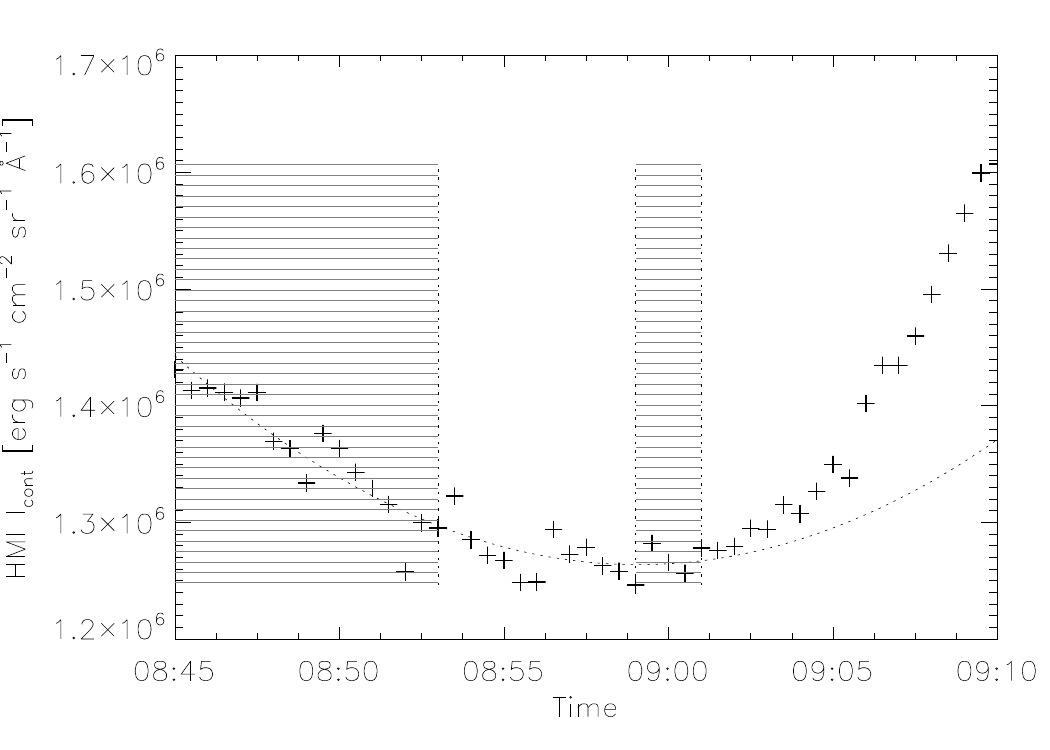 }\\
    \caption{Evolution of the signal in GREGOR/HiFI+ filters (upper panel, blue line indicates the blue continuum at 450.6~nm and red the TiO band) and the far wings of the \ion{Fe}{I} 6173\,\AA{} line from SDO/HMI (lower panel). The crosses indicate the measured signal in a pixel co-spatial with pixel No.\,53 on the IRIS slit. The dotted line indicates the fitted background, which represents the smooth evolution in the quieter period. The dashed regions indicate the time ranges used for the fitting of the smooth background.}
    \label{fig:optical_excess_px53}
\end{figure}

In our case, we first analysed the possible continuum enhancement in the broadband GREGOR/HiFI+ channels. We used only the channels of the blue continuum and TiO. We avoided the use of the G-band broadband filter because this spectral region is filled with spectral lines, and moreover there is a risk of cross-talk with the magnetic field. On the other hand, the lines in the TiO band appear only in sunspot umbrae because TiO molecules dissociate at $T > 4000$\,K in atmospheres of the Sun and cool stars \citep{Berdyugina2003, Bidaran2016}. Otherwise, the intensity observed in the TiO band corresponds to the continuum intensity. Since pixel No.\,53 is located at the edge of a pore where the probability of $T < 4000$\,K is low, we expected that the observed TiO band intensity is dominated by continuum and not by molecular lines. The procedure of converting the instrumental units to absolute units was somewhat complicated as compared to that described by \citet{Kleint_et_all_2016} due to the fact that the reference disc centre was out of the FOV of the HiFI+ observations, and hence one-to-one conversion could not be done directly. 

In order to overcome this issue, we first estimated the limb-darkening factor at the position of the IRIS slit. For the quiet-Sun reference, we chose pixels No.\,170--200 on the position of the IRIS slit and computed the median value for the blue continuum and TiO HiFI+ channels. At this position, the corresponding heliocentric distance is equal to $\mu=0.97$. Following the approximate formulae given by \cite{astrophysical_quantities_2000}, we determined limb-darkening factors of 0.984 for the blue continuum and 0.993 for TiO. By dividing the quiet-Sun reference on the slit by these factors, we obtained the expected value of the disc-centre intensity in the instrumental units. 

The disc-centre intensity in the absolute units was obtained by integrating the atlas spectrum \citep{1994svsp.coll...37N} with the known filter transmission profiles \citep[see Fig.~9 and Table~1 of][]{Denker23}. The resulting values were $0.417\times 10^7$~erg\,s$^{-1}$\,cm$^{-2}$\,sr$^{-1}$\,\AA$^{-1}$ for the blue continuum and $0.242\times 10^7$~erg\,s$^{-1}$\,cm$^{-2}$\,sr$^{-1}$\,\AA$^{-1}$ for TiO. The one-to-one comparison of the filter-integrated atlas spectrum in the absolute units with the estimate of the disc-centre intensity in the instrumental units obtained in the previous step led to the multiplicative conversion factors. These factors were then used to convert the intensity in the broadband filters at the position of the interesting pixel No.\,53 of the IRIS slit into absolute units. 

In this way, we obtained the time series of the intensity in pixel No.\,53 in the HiFI+ broadband channels. The time cadence of the HiFI+ imaging system is very short (on the order of milliseconds), so in order to minimise the effect of the atmospheric seeing, we averaged the consecutive frames from each burst of 500 images stored within one file by HiFI+. We averaged the 500 consecutive frames for TiO. The camera with the blue continuum filter took twice the time to record a burst; therefore, we divided each burst in half and averaged the corresponding consecutive 250 frames. The final averaged frames have a larger temporal cadence of about 5~s. We note that the blue continuum and TiO light curves in Fig.~\ref{fig:all_light_curves}e look flatter than in Fig.~\ref{fig:optical_excess_px53} due to the scaling and axis ranges.

As evident from the light curve of the GREGOR/HiFI+ measurements at a position equivalent to pixel No.\,53 on the IRIS slit (see Fig.~\ref{fig:optical_excess_px53} upper panel), there is no obvious continuum enhancement detected at the time of the flare except for an apparent brief enhancement after a short break in observations ($\sim$08:52\,UT), which was caused by bad seeing.  
The flare kernel is also not detectable in the running difference of the GREGOR/HiFI+ observations. Therefore, the enhancement, if any, has to be lower than the fluctuations of non-flare origin, such as seeing fluctuations. Hence, only an upper limit on the possible continuum excess may be determined. 

The upper limit on the excess was then determined as a root mean square value of the deviations of the signal from the parabolic fit in the period around the flare. The background fit was obtained while specifically excluding the span of the flare impulsive phase (see the dashed regions in Fig.~\ref{fig:optical_excess_px53}). In the physical units, those upper limits were 61~000~erg\,s$^{-1}$\,cm$^{-2}$\,sr$^{-1}$\,\AA$^{-1}$ for the blue continuum and 27~000~erg\,s$^{-1}$\,cm$^{-2}$\,sr$^{-1}$\,\AA$^{-1}$ for TiO.

For the HMI far-wing filtergrams, we followed the recipe by \cite{Kleint_et_all_2016}. The value of 45\,600~DN/s determined for the small box around the disc centre was compared to the expected atlas value of $0.315\times 10^7$~erg\,s$^{-1}$\,cm$^{-2}$\,sr$^{-1}$\,\AA$^{-1}$, and the ratio of the two served as a conversion factor from HMI instrumental values to absolute units. Once again, similar to the procedure performed for the GREGOR observations described above, we fitted a smooth background curve to the time series following exactly the same reasoning, and we computed the signal variations with respect to this smooth background change. In the corresponding light curve, no obvious enhancement at the time of the flare onset is visible. Hence (again, similar to the case of the GREGOR data), we estimated only the upper limit by calculating the root mean square values of the smooth-background residuals. The obtained value was about 270~DN/s, which corresponds to 19~000~erg\,s$^{-1}$\,cm$^{-2}$\,sr$^{-1}$\,\AA$^{-1}$ when assuming the conversion described above.

%________________________________________________________________
\section{Estimation of temperature in the flaring chromosphere}
\label{Sect:Temperature}

\subsection{Method}

Building on the method employed by \citet{Heinzel2017}, the temperature at the layer where the continuum enhancement is formed can be estimated. We assumed that any excess of the continuum emission is due to increased optically thin emission, $I_\nu$, from hydrogen recombination processes occurring in a heated layer of thickness~$D$:
\begin{equation}
I_\nu = n_{\mathrm{e}}^2 D\sum_{i}F_i(\nu, T)\,,
\end{equation}
where
\begin{equation}
F_i(\nu, T) \sim g_\mathrm{bf}(i,\nu) B_\nu(T) T^{-3/2} e^{h\nu_i/kT}(1-e^{-h\nu/kT}) / (i\nu)^3\,.
\end{equation}
There, $n_{\mathrm{e}}$ is the electron density, and the subscript $i$ (i\,=\,2,3,4,5) denotes different hydrogen recombination continua, Balmer, Paschen, Brackett, and Pfund, with corresponding wavelengths at continuum heads of $\lambda_2 = 3646\,\mbox{\AA}$, $\lambda_3 = 8204\,\mbox{\AA}$, $\lambda_4 = 14584\,\mbox{\AA}$, $\lambda_5 = 22790\,\mbox{\AA}$, respectively. The Gaunt factor is $g_\mathrm{bf}(i,\nu)\approx 1$. The term $n_{\mathrm{e}}^2 F_i(\nu, T)$ is the recombination emissivity \citep[see, e.g.,][]{Heinzel2017}, and $n_{\mathrm{e}}^2D$ is the emission measure \citep[see, e.g.,][]{Dominique2018}.

This method requires two different wavelengths in the optical (Opt) and in the UV range. The bound-free emission at those wavelengths is

\begin{eqnarray}
\label{eq:continua}
    \nonumber I_\mathrm{UV}&=&n_{\mathrm{e}}^2 D\sum_{i=2}^{5}F_i(\nu_\mathrm{UV}, T)\,,\\%\qquad
    I_\mathrm{Opt}&=&n_{\mathrm{e}}^2 D\sum_{i=3}^{5}F_i(\nu_\mathrm{Opt}, T)\,.
    \end{eqnarray}

Assuming that all recombination continua are formed within the same layer, the ratio of their intensities is then only a function of temperature,

\begin{equation}\label{eq:ratio}
        I_\mathrm{Opt}/I_\mathrm{UV} = f(T) \, .
\end{equation}

We point out that although the ratio is dimensionless, its numerical values depend on whether $I_\nu$ or $I_\lambda$ is used. Here, we used $I_\lambda$, as all observed intensities were calibrated per wavelength. Additionally, we note that \citet{Kerr2014} used a similar ratio method for their optically thick black-body model of flare white-light emission. Likewise, our approach follows similar steps as their other method to constrain the temperature of a chromospheric slab under an optically thin assumption.

\begin{figure}
\centering
\includegraphics[width=1\linewidth]{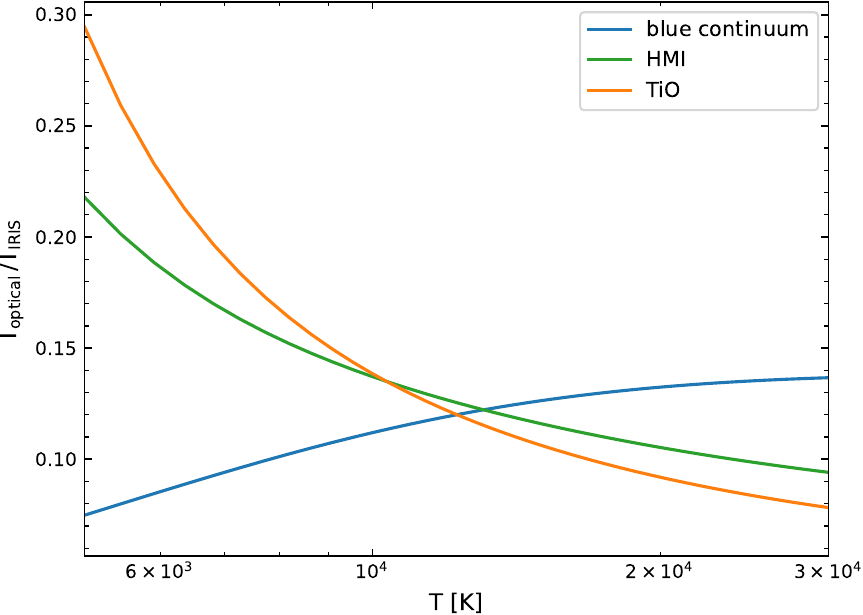}
\caption{Theoretical ratios, $f(T)$, for three wavelengths at the optical range: $\lambda_{\mbox{blue}}$, $\lambda_{\mbox{HMI}}$, and $\lambda_{\mbox{TiO}}$.}
\label{fig:bfratios}
\end{figure}

This approach is straightforward to apply to off-limb flares, for example those in \citet{Heinzel2017}, where $D$ is the line of sight geometrical thickness of a flare loop. For such cases, it is reasonable to assume optically thin emission. The stratification of electron density and temperature along an off-limb flare loop then results in continuum intensity stratification with height.  

In our case, that is for a flare on the disc, the observed intensity is given by the integral of the emissivity contributions along the flare loop. Therefore, the excess of intensity, $\Delta I_\nu$, with respect to a preflare state is
\begin{equation}
    \Delta I_\nu \approx \int n_{\mathrm{e}}^2(z)\sum_i F_i(\nu, T(z)) e^{-\tau(z)} \mathrm{d}z \approx D \left< n_{\mathrm{e}}^2 \right> \left<\sum_i F_i(\nu, T)\right>,
\end{equation}
still assuming that the flare intensity enhancement is optically thin and that the photospheric layers where the optically thick part of the continuum emission is formed are not affected by the flare process. Therefore, the contribution of the undisturbed photosphere can be subtracted. This approach agrees with \citet{Potts2010ApJ}, who showed that optical continuum emission detected in X-class flare ribbons observed on the solar disc is of small optical depth.

Furthermore, assuming that the flare intensity enhancement comes predominantly from similar layers for both chosen wavelengths, the mean value $\left< n_{\mathrm{e}}^2 \right>$ cancels out, and the enhancement intensity ratio depends on the temperature in a manner similar to the way it does in Eq.~(\ref{eq:ratio}):

        \begin{equation}\label{eq:ratio_enh}
        \Delta I_\mathrm{Opt}/\Delta I_\mathrm{UV} = f\left(\left<T\right>\right) \,,
    \end{equation}
where $\left<T\right>$ is the mean temperature of the layers where the continuum enhancement originates.

Following the conclusions of \citet{Heinzel2017}, this analysis neglected contributions from the hydrogen free-free emission and Thompson scattering. Details of the method and testing of its validity are beyond the scope of this work, but they are subjects of a follow-up paper \citep{Kasparova-prep}.

\subsection{Determination of \texorpdfstring{$\left<T\right>$}{TEXT} }

\begin{figure}
\centering
\includegraphics[width=1\linewidth]{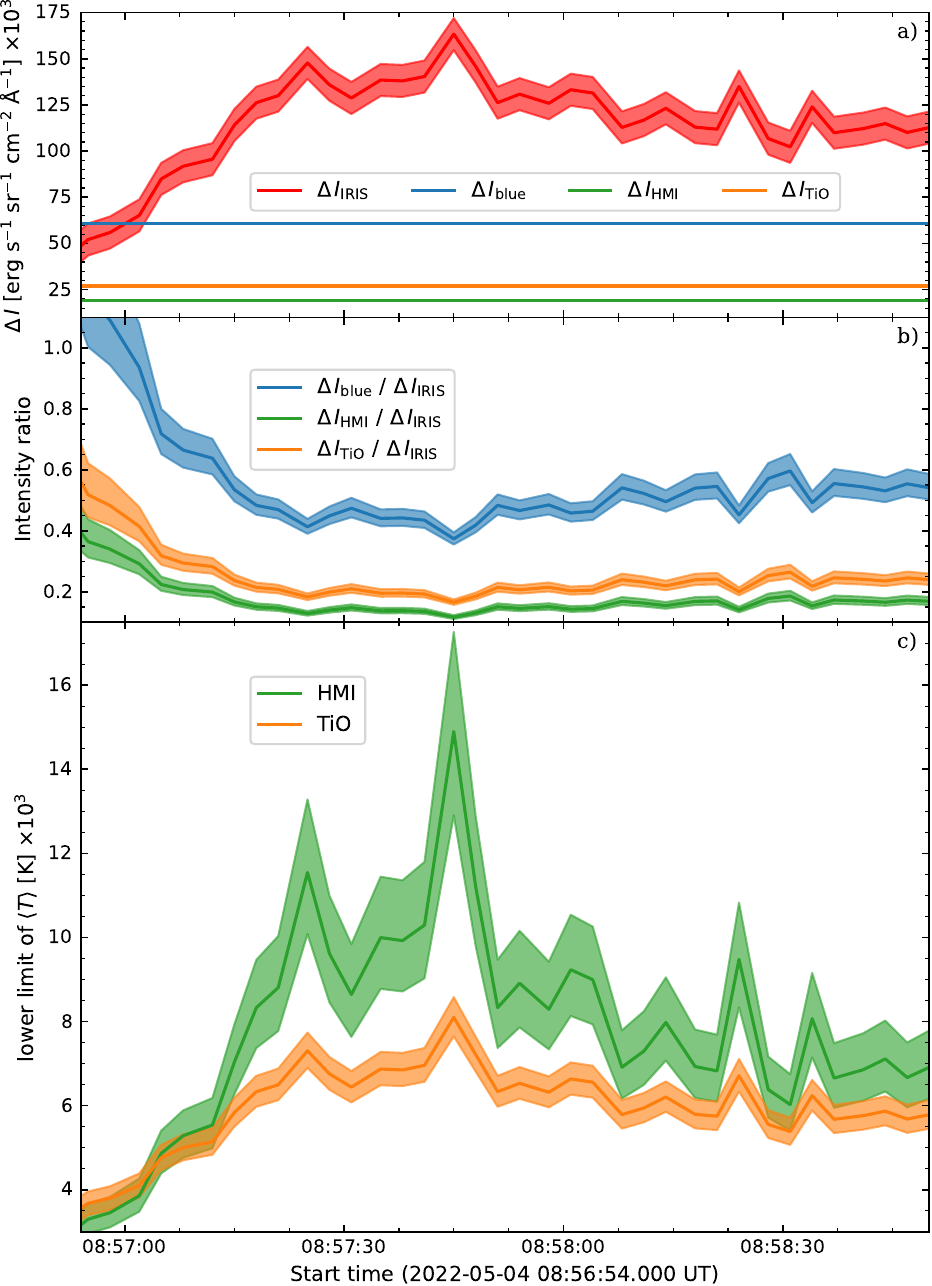}
\caption{Time evolution of (a) the determined intensity enhancement at $\lambda_{\mbox{IRIS}}$ and the upper limits at $\lambda_{\mbox{blue}}$, $\lambda_{\mbox{HMI}}$, and $\lambda_{\mbox{TiO}}$; (b) the measured ratios for the three different wavelengths at the optical range: $\lambda_{\mbox{blue}}$, $\lambda_{\mbox{HMI}}$, and $\lambda_{\mbox{TiO}}$; (c) the lower limit on the mean temperature as determined from $\Delta I_\mathrm{TiO}/\Delta I_\mathrm{IRIS}$ and $\Delta I_\mathrm{HMI}/\Delta I_\mathrm{IRIS}$.}
\label{fig:temp_lc}
\end{figure}

This rather simple method permits an estimation of $\left<T\right>$ by solving Eq.~(\ref{eq:ratio_enh}) for the measured enhancement ratios. The temperature dependence of the theoretical enhancement ratios at typical flare chromospheric values is shown in Fig.~\ref{fig:bfratios}. In evaluating function $f$, Eq.~(\ref{eq:ratio_enh}), we accounted for relevant hydrogen recombination continua as in Eq.~(\ref{eq:continua}). We calculated the ratio between optical (HMI and TiO) and NUV (IRIS band C) enhancements, as seen in Fig.~\ref{fig:temp_lc}b. For this calculation, we used the enhancement of the IRIS band C above the preflare level and the determined upper limits of enhancement for HMI and TiO, estimated in Sec.~\ref{Sect:IRIS_continuum} and Sec.~\ref{Sect:optical_continuum}, respectively, and summarised in Fig.~\ref{fig:temp_lc}a.  

The resulting time evolution of $\left<T\right>$ obtained from two enhancement ratios, namely $\Delta I_\mathrm{HMI}/\Delta I_\mathrm{IRIS}$ and $\Delta I_\mathrm{TiO}/\Delta I_\mathrm{IRIS}$, is plotted in Fig.~\ref{fig:temp_lc}c. The time interval is constrained to the period of the first intensity increase in IRIS band C. The shaded region indicates the spread of $\left<T\right>$ values, taking into account the standard deviation of the intensities in band C during the preflare interval (see Sec.~\ref{Sect:IRIS_continuum}). We note that since we used upper limits of HMI and TiO enhancements, we determined lower limits of the mean temperature $\left<T\right>$. The real optical continuum enhancements could be lower and therefore lead to smaller ratios and higher temperatures, according to the theoretical behaviour of $f(T)$ for HMI and TiO wavelengths in Fig.~\ref{fig:bfratios}. Also, we point out that the time resolution of the temperature light curve is dictated by the time resolution of the IRIS spectral data.

The HiFI+ blue filter data and the corresponding $\Delta I_\mathrm{blue}/\Delta I_\mathrm{IRIS}$ ratio were not used for the determination of $\left<T\right>$ because the limit of enhancements in the GREGOR blue filter (Sec.~\ref{Sect:optical_continuum}) leads to ratios higher than the theoretical ratios (the blue curve in Fig.~\ref{fig:bfratios} is lower than $\sim 0.14$, and its behaviour does not change for higher temperatures). Similarly, as for the limits of HMI and TiO enhancements, the real enhancement in the blue filter could be lower but not zero. Our approach to determine the temperature of the layers where the flare continuum emission originates is valid only for non-zero optical enhancements. If we assume that the detected enhancement in the IRIS NUV continuum is due to the bound-free contribution, there must be a corresponding contribution in the optical continuum. Yet, it could be below the detection limits of the data we have at our disposal.

Additionally, the mean electron density, $\left< n_\mathrm{e} \right>$, could be determined from the observed enhancement at IRIS band C and the modelled enhancement evaluated for the obtained $\left< T \right>$ (see Eq.~\ref{eq:continua}). That is $\left< n_\mathrm{e} \right> = \left(\Delta I_\mathrm{IRIS} / I_\mathrm{IRIS} / D\right)^{1/2}$. The thickness of the layer where the continuum enhancements are formed, $D$, was obtained from \citet[][Figs. 1, 2; Sec. 4.1]{Heinzel2017}. In that work, the optical continuum enhancements in the analysed off-limb flares extend over $\sim 500$~km above the limb, as seen in the HMI data. Adopting the results presented above, we obtained $\left< n_\mathrm{e} \right> \sim 1\times10^{13}$~cm${}^{-3}$.

\section{Discussion}
\label{Sect:Discussion}

Based on \citet{Heinze2014}, the observed NUV flare continuum emission is produced by bound-free transitions, which implies that there is an optical continuum enhancement as well. Due to the limitations to detect the continuum by HMI and to the seeing-induced intensity fluctuations at GREGOR, the flare kernel was not unambiguously detected in the optical broadband observations. However, the construction of the light curves at the location corresponding to pixel No.\,53 on IRIS slit allowed us to estimate an upper limit on the brightness changes caused by the flare. These upper limits contributed to constraining of the temperature of the heated atmosphere.

We determined a lower limit on the mean temperature and its time evolution in the layer where the optically thin flare continuum enhancement is formed without assuming any value for the so-called emission measure, $n_{\mathrm{e}}^2 D$ -- in contrast with previous similar works (see, e.g., \citet{Heinzel2017, Kerr2014, Dominique2018}). The determined lower limits of the temperature vary in the $\sim(3 - 15)\times 10^3$~K range, which is similar to the temperatures obtained in the radiation-hydrodynamical simulations of beam heating of an initial VAL C-like atmosphere \citep{Heinzel2017, Kasparova2019, carlsson2023} and to values determined by \cite{Kerr2014} in their optically thin scenario. 

Additionally, we obtained the mean electron density of the emitting chromospheric layer, $\left< n_\mathrm{e} \right> \sim 1\times10^{13}$~cm${}^{-3}$, which is similar to the modelled values in \citet{Heinzel2017} and those discussed in \citet{Kerr2014,Dominique2018}. It also lies within the transition region densities determined from IRIS FUV line ratios \citep{Polito2016b, Joshi2021b} or obtained in numerical models \citep{Kerr2019}.

The time evolution of the lower limit of the temperature shows rapid variations on a timescale of several seconds (see Fig.~\ref{fig:temp_lc}), reflecting the IRIS temporal resolution of $\sim 3$~s. Individual peaks last less than $\sim$10\,s, with a single-point maximum, meaning that they are likely strongly undersampled, and possibly have an even shorter duration. We note that similar fast changes in the intensity of different flaring lines have been observed before \citep[see, e.g.,][]{Jeffrey18,Lorincik22b,Lorincik22a}. The exact nature of these variations is beyond the scope of our present paper; nevertheless, there are several conceivable mechanisms, including intermittent particle acceleration (as suggested by both the GOES derivative and the multi-peak Konus-Wind HXR light curve) and turbulence \citep{Jeffrey18}, which itself can be related to particle acceleration \citep{Bian14}. 
Finally, we note that microwave emission in the ~5-15 GHz range shows similar time evolution to the HXR light curve \citep{2023_4MayFlare_smirnova}. This suggests that both types of emission were produced by the same population of non-thermal elections.

Using the values of $\left< T \right>$, $\left< n_\mathrm{e} \right>$, and the limb-darkening factor for near-H$\alpha$ continuum, we estimated the expected relative continuum enhancement for FICUS. Since 15\% of the observed area was covered by the H$\alpha$ solar flare (based on Fig.~\ref{Fig:Overview}m), the relative continuum enhancement was below 1\%. Such an enhancement is smaller than the FICUS data uncertainties for this particular flare and confirms the lack of enhancement detection in FICUS near-H$\alpha$ continuum.

\section{Summary}
\label{Sect:Summary}

We have presented the first results of the first successful joint GREGOR and IRIS observing campaign of solar flares. The campaign was supported by ground-based instruments located at the Astronomical Institute in Ond\v rejov. On 2022 May 4, we observed a confined M5.7 solar flare, with one flare ribbon crossing the slit of the IRIS spectrograph. We investigated the temporal evolution of spectral (IRIS NUV) and imaging (GREGOR HiFI+) data in a rather faint flare kernel at the edge of a pore. These data were complemented by space-based observations from AIA, HMI, GOES-16, and Konus.

We focused on the observations of the NUV and optical flare continuum obtained with unprecedented temporal resolution by IRIS and HiFI+, respectively. The NUV spectra recorded at full read-out mode allowed us to examine continuum enhancements at wavelengths farther away from the strong Mg II lines than in previous studies. We found that the NUV continuum emission showed rapid temporal variations. On the other hand, due to limitations of the HMI instrument and seeing conditions at GREGOR, the HiFI+ and HMI datasets allowed us to estimate only an upper limit of the flare continuum enhancement in the optical range. The spatially integrated HXR light curves supported our interpretation of flare-related continuum emission.

%The spectral distance from the line wings is especially important for stronger flares. Sit-and-stare 

The combination of the optical and NUV continuum analyses led to a lower limit on the temperature of $\sim(3 - 15)\times 10^3$~K %3--15 $\times 10^3$\,K 
and a mean electron density of $\sim 1\times10^{13}$~cm${}^{-3}$ in the flaring chromospheric layers. This study demonstrates the rich diagnostic potential that multi-instrument and multi-wavelength flare campaigns have. Our findings support previous indications that continuum enhancement is also present in confined flares and in flaring areas of weak emission and that the associated temperature varies on very short timescales.

\begin{acknowledgements}
A.B., J.D., J.K., M.G.R., M.P., M.S., M.\v{S}., M.Z., and W.L. acknowledge the institutional support RVO:67985815 from the Czech Academy of Sciences. J.D. and J.K. acknowledge the Czech National Science Foundation, Grant No. GACR 22-07155S, and G.G.M.,  M.G.R., and M.Z. acknowledge Grant No. GACR 21-16508J. A.B. also acknowledges the support within the ”Excellence Initiative-Research University” for years 2020-2026 at University of Wrocław, project no. BPIDUB.4610.96.2021.KG. and BPIDUB.4610.15.2021.KP.B.
We thank Alexandra Lysenko for providing us with Konus-Wind data, \href{http://www.ioffe.ru/LEA/kwsun/}{ http://www.ioffe.ru/LEA/kwsun/}. This research data leading to the results obtained has been supported by SOLARNET project that has received funding from the European Union’s Horizon 2020 research and innovation programme under grant agreement No. 824135. The 1.5-metre GREGOR solar telescope was built by a German consortium under the leadership of the Institute for Solar Physics (KIS) in Freiburg with the Leibniz Institute for Astrophysics Potsdam, the Institute for Astrophysics G\"ottingen, and the Max Planck Institute for Solar System Research in G\"ottingen as partners, and with contributions by the Instituto de Astrof\'isica de Canarias and the Astronomical Institute of the Academy of Sciences of the Czech Republic. IRIS is a NASA small explorer mission developed and operated by LMSAL with mission operations executed at NASA Ames Research Center and major contributions to downlink communications funded by ESA and the Norwegian Space Centre.
\end{acknowledgements}

\bibliographystyle{aa}
\bibliography{2022-05-04_Flare_GREGOR_IRIS}

\begin{appendix} %First appendix

\section{IRIS pointing wobble during the 2022 May 4 flare}
\label{Sect:Appendix_wobble}

\begin{figure}
\centering
\includegraphics[width=8.8cm, viewport= 0 65 997 440, clip]{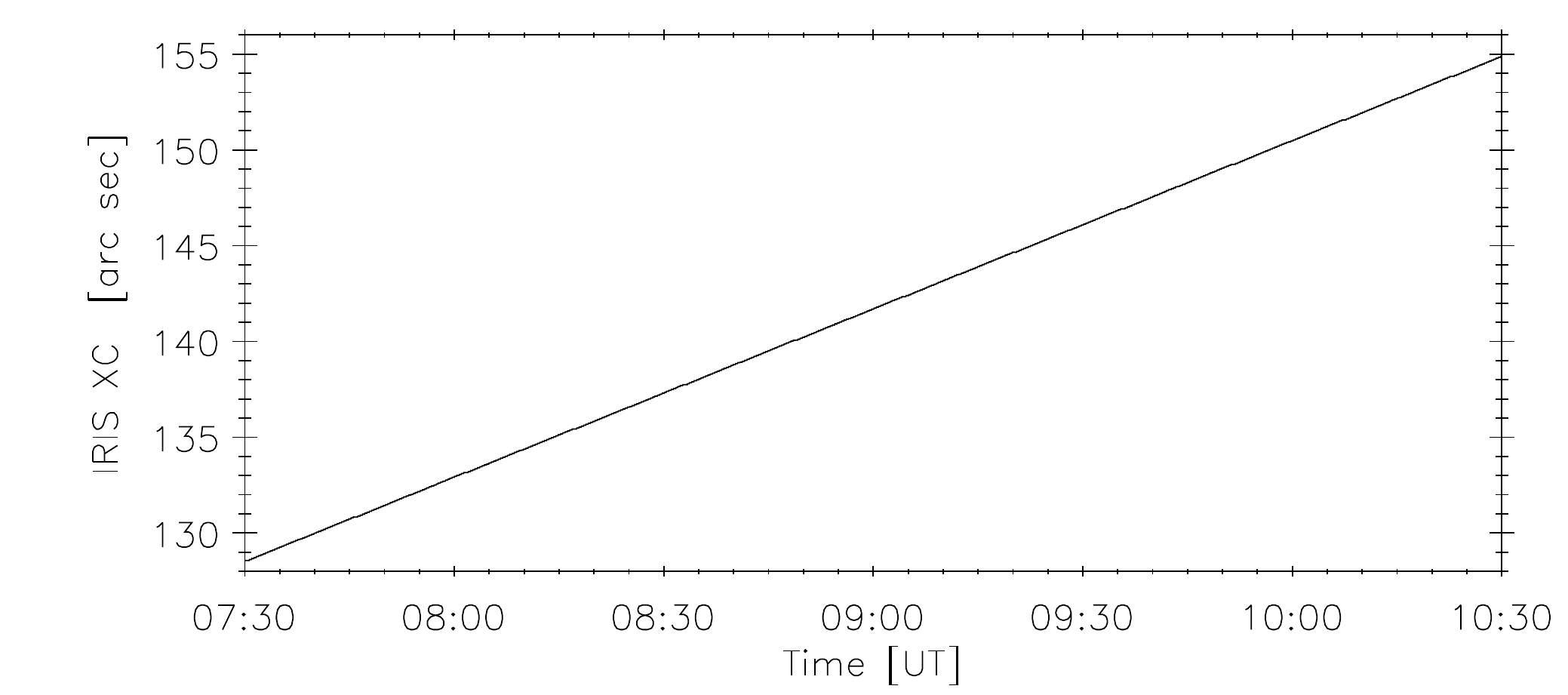 }
\includegraphics[width=8.8cm, viewport= 0  0 997 440, clip]{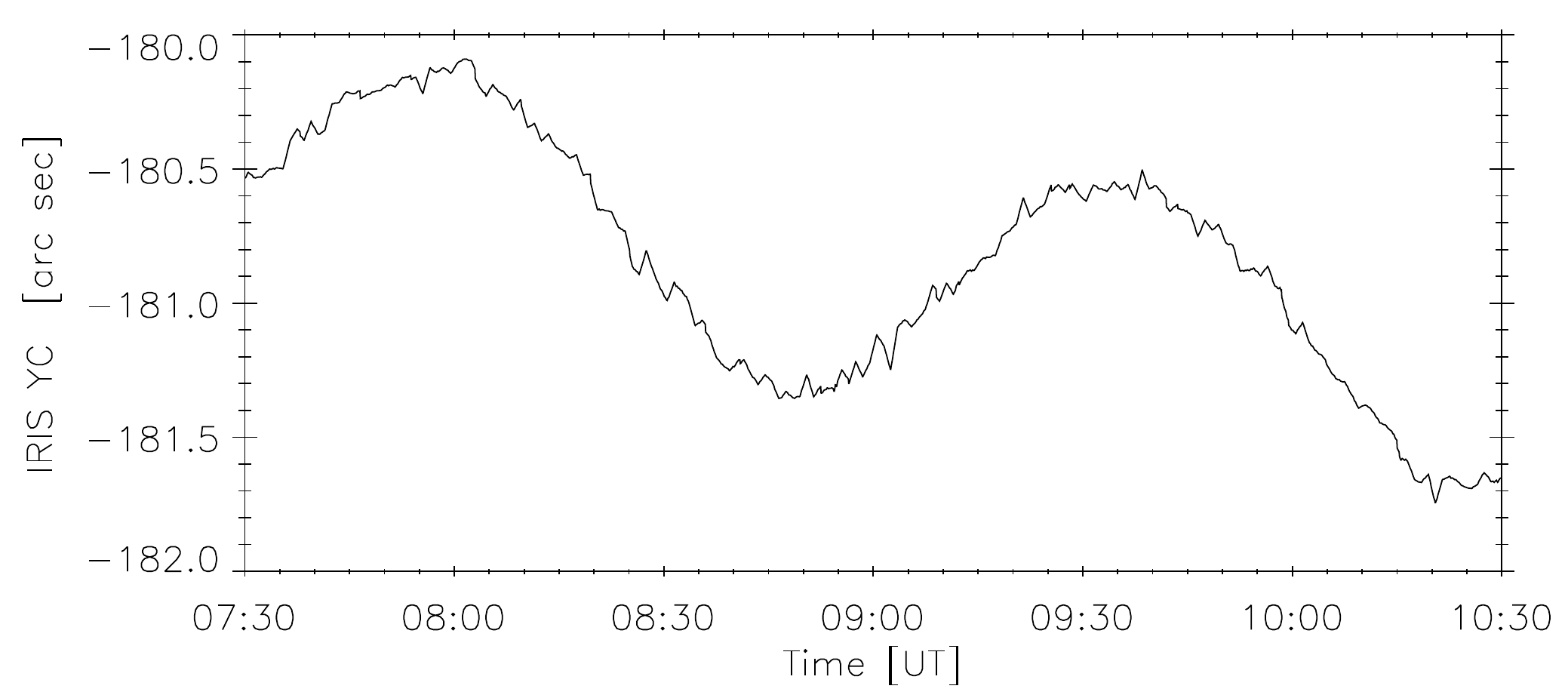 }
\caption{IRIS pointing during the HOP\,422 on 2022 May 4 showing the evolution of the pointing in terms of the centre of the SJI FOV [$XC$, $YC$].}
\label{Fig:Appendix_IRIS_pointing}
\end{figure}

The IRIS light curve analysis presented in Sect. \ref{Sect:IRIS_continuum} is complicated by the changes in the IRIS pointing. During the HOP\,422 (OBS 3884855852: Medium sit-and-stare), lasting from 7:30 to about 13:00\,UT on 2022 May 4, IRIS tracked the AR 13004. This tracking is clearly identifiable as a continuous increase of the IRIS $XC$, the Solar $X$ coordinate of the SJI FOV (top panel of Fig.~\ref{Fig:Appendix_IRIS_pointing}). Except minute residuals, this tracking is easily rectified with the correction of the solar differential rotation.

However, the pointing in $Y$ was not stable, as indicated by the IRIS $YC$ coordinate. Here, IRIS showed an orbital wobble with an amplitude of about $0.6\arcsec$; that is, about 2 pixels (in the spatially binned SJI images). The wobble, as indicated by the bottom panel of Fig.~\ref{Fig:Appendix_IRIS_pointing}, is primarily sinusoidal, but with a secular component. This secular component is reduced, but not removed, by the correction for differential rotation. The wobble is not affected by this procedure.

We note that the orbital wobble occurs due to thermal bending of the mounting of the IRIS guide telescope \citep[see Sect. 4.5 of][]{DePontieu14} and is in phase with the IRIS orientation with respect to the illuminated side of the Earth. The orbital wobble is normally corrected using orbital wobble tables\footnote{\url{https://iris.lmsal.com/itn51/iris_timeline.html}} to less than about $0.3\arcsec$ \citep{DePontieu14}. Why the wobble is larger in the present observation is not known to us.

\section{Testing the absolute calibration of IRIS near-ultraviolet spectra}
\label{Sect:Appendix_iris_calib}

\begin{figure}
\centering
\includegraphics[width=8.8cm]{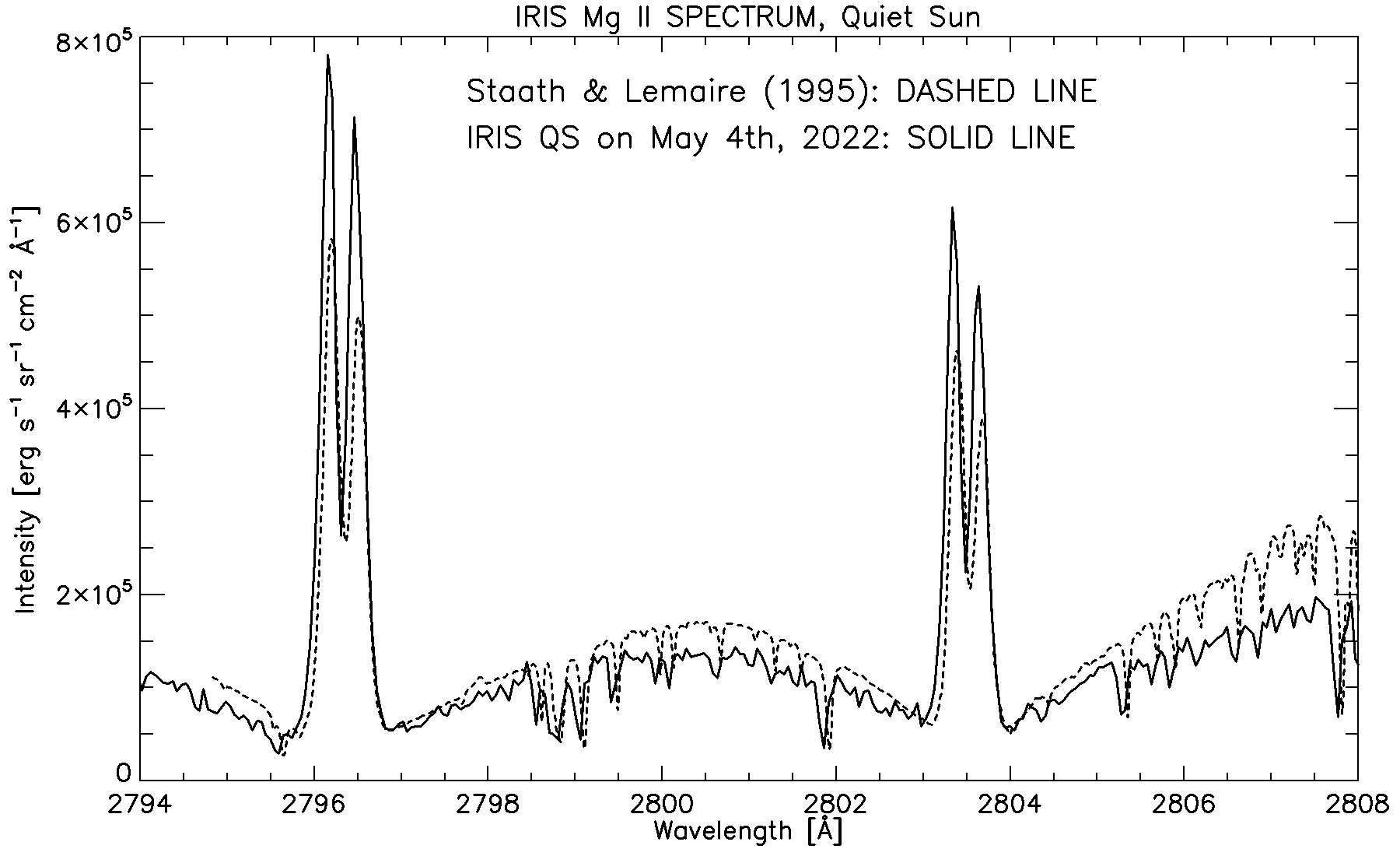}
\caption{Comparison of our calibrated mean quiet-Sun disc spectrum obtained on 2022 May 4 with RASOLBA calibrated spectrum.}
\label{Fig:Appendix_IRIS_calibration}
\end{figure}

In order to verify the IRIS calibration procedure and to check if it gives a reasonable results we compared the \ion{Mg}{II} quiet-Sun line profiles observed with IRIS slit with the previous calibrated observations obtained with RASOLBA experiment \citep{Staath95}. For this purpose, we chose a pixel (pixel No.\,190 along IRIS slit) as distant as possible from the observed AR and generated a reference QS spectrum. The results of comparing our calibrated QS spectrum with RASOLBA calibration is presented in Fig.~\ref{Fig:Appendix_IRIS_calibration}. The good agreement between both spectra confirms the correctness of our radiometric calibration. Some differences of the signal, in particular in the line wings, could be caused by heterogeneity of the photospheric brightness.

\end{appendix}

\end{document}